\documentclass[twocolumn,preprintnumbers,amsmath,amssymb,amsfonts,final,aps,citeautoscript,footinbib,prl,superscriptaddress,longbibliography,10pt]{revtex4-2}
\usepackage{graphicx}
\usepackage{tabularx} 
\usepackage[caption=false]{subfig} 
\usepackage{amsmath,amssymb,amsfonts,pifont,fancybox,float,mathtools}
\usepackage{graphicx}
\usepackage{epsfig}
\usepackage{braket}
\usepackage{kantlipsum}
\usepackage[normalem]{ulem}
\usepackage{color}
\usepackage[pdftex,dvipsnames]{xcolor}  
\usepackage[colorlinks,bookmarks=true,citecolor=blue,linkcolor=blue,urlcolor=blue, breaklinks=true]{hyperref}
\usepackage{simplewick}
\usepackage[latin1]{inputenc}
\usepackage{mathrsfs}
\usepackage[vcentermath,noautoscale]{youngtab}

\newcommand{\be}{\text{e}}

\usepackage{epstopdf}
\newcommand{\para}[1]{\noindent\textbf{\textit{#1}.}\hspace{0.5cm}}
\newcommand{\SU}{\ensuremath{\mathrm{SU}}}
\newcommand{\SUN}{\ensuremath{\mathrm{SU}(N)}}
\newcommand{\PSU}{\ensuremath{\mathrm{PSU}}}

\newcommand{\Z}{\ensuremath{\mathbb{Z}}}
\newcommand{\three}{\ensuremath{\mathbf{3}}}
\newcommand{\threeb}{\ensuremath{\mathbf{\bar{3}}}}
\newcommand{\eight}{\ensuremath{\mathbf{8}}}
\newcommand{\class}[1]{\ensuremath{\lfloor #1 \rfloor}}

\graphicspath{{./figures/}{./Figs/}}

\begin{document}

\title{Non-Landau quantum phase transition in modulated SU($N$) Heisenberg spin chains}

\author{Sylvain Capponi} 
\affiliation{Laboratoire de Physique Th\'eorique, Universit\'e de Toulouse, CNRS, UPS, France.}

\author{Lukas Devos}
\affiliation{Department of Physics and Astronomy, Ghent University, Belgium}

\author{Philippe Lecheminant} 
\affiliation{Laboratoire de Physique Th\'eorique et Mod\'elisation, CNRS, CY Cergy Paris Universit\'e, 95302 Cergy-Pontoise Cedex, France.}

\author{Keisuke Totsuka} 
\affiliation{Center for Gravitational Physics and Quantum Information, 
Yukawa Institute for Theoretical Physics, Kyoto University, Kyoto 606-8502, Japan.}

\author{Laurens Vanderstraeten}
\affiliation{Center for Nonlinear Phenomena and Complex Systems, Universit\'e Libre de Bruxelles, Belgium}

\date{\today}

\begin{abstract}

We investigate the nature of the quantum phase transition in modulated $\SUN$ Heisenberg spin chains. In the odd-$N$ case, the transition separates a trivial non-degenerate phase to a doubly-degenerate gapped chiral $\PSU(N)$ symmetry-protected topological (SPT) phase which breaks spontaneously the inversion symmetry. The transition is not an Ising transition associated to the breaking of the $\Z_2$ inversion symmetry, but is governed by the delocalization of the edge states of the SPT phase. In this respect, a modulated $\SU(N)$ Heisenberg spin chain provides a simple example in one dimension of a non-Landau phase transition which is described by the $\SU(N)_1$ conformal field theory. We show that the chiral SPT phase exhibits fractionalized spinon excitations, which can be confined by changing the model parameters slightly.
\end{abstract}
\maketitle

\begin{figure}
    \centering
    \includegraphics[width=\columnwidth]{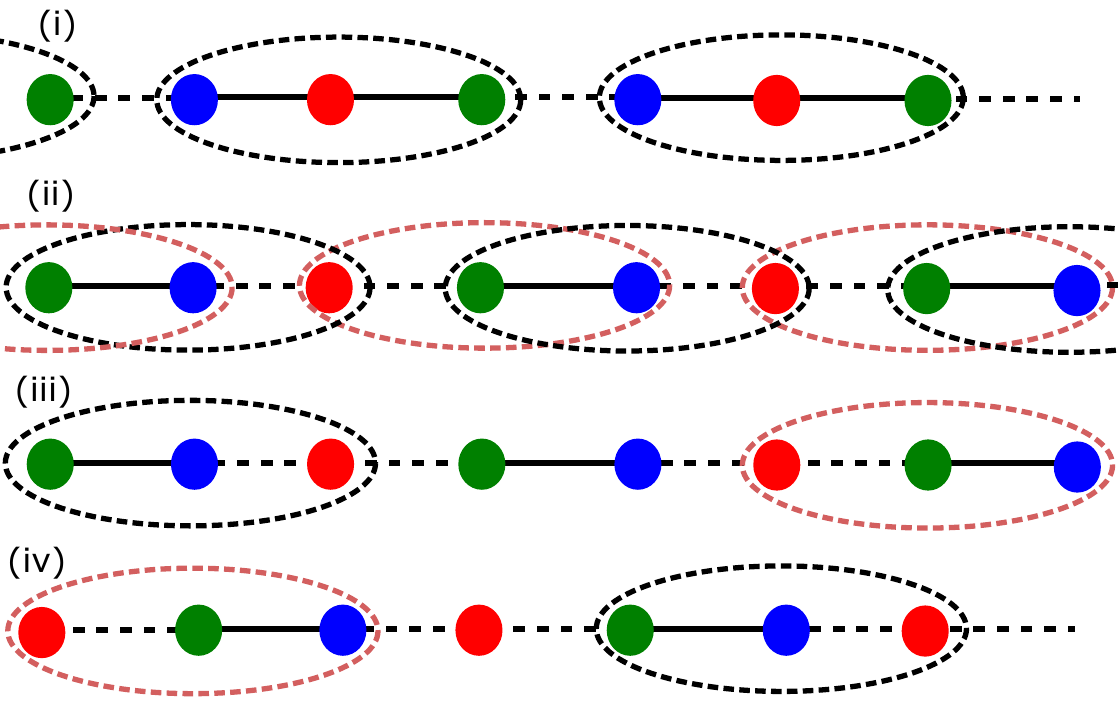}
    \caption{Cartoon pictures of the SU($3$) ground states where strong/weak antiferromagnetic couplings are depicted as full/dashed bonds respectively. From top to bottom: (i) $\delta<0$ (connected to a trivial product of singlets); (ii) $\delta>0$ (two possible ground states, related by inversion symmetry, are shown with different dashed ovals); (iii)-(iv) two soliton excitations (respectively $\threeb$ and $\three$) that can exist as domain walls between the two possible ground states for $\delta>0$.}
    \label{fig:su3}
\end{figure}

\para{Introduction}%
The Landau-Ginzburg-Wilson (LGW) paradigm provides the conceptual framework to describe continuous phase transitions in modern condensed matter \cite{Herbut-book-07}. The central idea of the approach is that the universal properties of a transition are fully characterized by the long-wavelength, long-time fluctuations of a symmetry-breaking order parameter. In the past three decades, however, many exotic quantum phase transitions beyond the LGW paradigm have been proposed. A prime example is the continuous phase transition between two phases with the same symmetry but with different topological orders \cite{Wen-W-93,Wen-20}. The underlying transition is not described by the fluctuations of a Landau order parameter but stems from a drastic change in the long-range quantum entanglement pattern of the underlying topological orders of the phases \cite{Wen-RMP-17}.
 
A particularly striking example of a non-Landau transition arises in the concept of deconfined quantum criticality (DQC), with a possible direct continuous transition between two phases with incompatible spontaneous broken symmetries as for the N\'eel to valence-bond-solid transition of two-dimensional competing spin-1/2 magnets \cite{Senthil-V-B-S-F-04,Senthil-B-S-V-F-04,Senthil-23}. The transition is described by emergent deconfined gauge fields coupled to fractionalized degrees of freedom, whereas they are confined in the conventional phases on either side of the transition \cite{Senthil-23}. This Landau-forbidden transition has attracted much interest over the years and has recently become very relevant with its possible experimental observation in a pressurized SrCu$_2$(BO$_3$)$_2$ compound \cite{Zayed2017,Guo2020,Cui2023,Guo2023}. In one dimension (1D), there are several examples of non-Landau continuous phase transitions, some of them realizing a 1D-version of a DQC point \cite{Tsui2017,Mudry-F-M-T-H-19,Jiang-M-19,Roberts-J-M-19, Huang-LYMX-19, Huang-Y-20,Yang-Y-S-20, Brenden-M-21,Zheng-S-L-22,Zhang-Levin-23,Yang-P-L-Y-23,Lee-R-M-V-H-C-23,Romen-B-K-24}.  
These models are defined from discrete or U(1) symmetries with a U(1) Luttinger criticality at the transition \cite{Haldane-JPC-81}.

In this Letter, we introduce a general class of 1D lattice models with non-abelian SU($N$) continuous symmetry which display non-Landau quantum phase transitions. We consider an SU($N$) Heisenberg spin chain with an explicit modulation of the interactions with period $N$: 
\begin{equation}
 H_{\delta} =    J \sum_{i=1}^{NL} \sum_{A=1}^{N^2-1}
 \left\{ 1 + \delta \cos \left( \frac{2 \pi i}{N}  \right) \right\} S^{A}_{i}  S^{A}_{i+1},
\label{eq:hamhidaNodd}
\end{equation}
$S^{A}_{i}$ being the SU($N$) spin operators on the $i$-th site of the chain which transform in the $N$-dimensional fundamental representation of the $\SU(N)$ group, normalized as $\text{Tr}(S^{A} S^{B})=\delta^{AB}/2$. In the simplest $N=2$ case, the model corresponds to the alternating spin-1/2 Heisenberg spin chain with explicit dimerization which was introduced by Hida in Ref.~\onlinecite{Hida-92a} to describe the main properties of the Haldane phase  of the spin-1 Heisenberg chain, the paradigmatic example of a 1D interacting symmetry-protected topological (SPT) phase \cite{Haldane-Novel-lec-17}.

In this work, we map out the phase diagram of model (\ref{eq:hamhidaNodd}) for general odd $N$ at zero temperature by means of complementary non-abelian bosonization and numerical approaches. The phase transition is always located at $\delta = 0$ and belongs to the $\SU(N)_1$ universality class with central charge $c=N-1$. It is shown that the continuous transition for odd $N$ represents a simple example of a non-Landau transition in 1D, similarly to the result that was obtained for $N=3$ in a related model in Ref.~\onlinecite{Bi-L-S-20}. The quantum critical point separates a trivial phase at $\delta <0$ from a twofold-degenerate gapped phase at $\delta >0$. The latter exhibits two chiral-SPT ground states that are protected by the projective unitary symmetry $\PSU(N)=\SU(N)/\Z_N$ \cite{Duivenvoorden-Q-13}, for which the inversion symmetry $\mathcal{I}$ ($S^{A}_{NL-n} \rightarrow S^{A}_{n+1}$) of the model is spontaneously broken --  see Fig.~\ref{fig:su3} for an illustration of the ground state in the simplest $N=3$ case. Despite this symmetry breaking, the transition is \emph{not} an Ising transition described by the fluctuations of a ${\mathbb{Z}}_2$ order parameter related to the inversion symmetry breaking. In contrast, the $\SU(N)_1$ quantum criticality of the transition stems from the delocalization of the edge states of the two degenerate chiral-SPT ground states which are exchanged under the inversion symmetry.

\para{Weak-coupling approach}%
The continuum limit of model (\ref{eq:hamhidaNodd}) is performed by exploiting the fact that the low-energy properties of the uniform SU($N$) Heisenberg spin chain for $\delta=0$ is described by an $\SU(N)_1$ conformal field theory (CFT) 
\cite{Affleck-NP86, Affleck-88, James-K-L-R-T-18}. In the low-energy limit, the lattice spin operators are described by: 
\begin{equation} \label{eq:spinop}
S^{A}_{n}/a_0 \simeq  J^{A}_{\text{L}} +  J^{A}_{\text{R}} 
+  i  \lambda \, \be^{\frac{ i 2 \pi }{Na_0}x} \; {\rm Tr} ( g (x)  T^A) + \text{H.c.},
\end{equation}
where $x=n a_0$, $a_0$ being the lattice spacing, and 
$\lambda = Ce^{ i \theta_0}$ ($C >0$) is a non-universal complex constant.
In Eq. (\ref{eq:spinop}), $J^{A}_{\text{R,L}}$ are the chiral $\SU(N)_1$ currents which generate the $\SU(N)_1$ CFT, $g$ is the $\SU(N)_{1}$ primary field with scaling dimension $(N-1)/N$, and $T^A$ are the SU($N$) generators in the fundamental representation of the $\SU(N)$ group. Two important discrete lattice symmetries are the one-step translation symmetry $\mathcal{T}_{a_0}$ which is explicitly broken when $\delta \ne 0$ and the inversion symmetry $\mathcal{I}$ which is always a symmetry. Using the correspondence \eqref{eq:spinop}, these two symmetries are implemented in the low-energy approach by the identification:
\begin{equation}
\begin{split}
g \xrightarrow{\mathcal{T}_{a_0}} &   \;  \mathrm{e}^{i \frac{2 \pi}{N}} g   \\
g(x) \xrightarrow{\mathcal{I}} &   - \mathrm{e}^{ - 2 i \theta_0 }  \mathrm{e}^{-i \frac{2 \pi}{N}} g^{\dagger}(-x)  .
\end{split}
\label{eq:contlimitsym}
\end{equation}
When $|\delta| \ll 1$, a low-energy approach for the spin-chain model can be derived by means of the identification (\ref{eq:spinop}). Its Hamiltonian density reads as follows:
\begin{equation}
\begin{split}
 {\cal H}_{\delta} =&    {\cal H}_{0} +  {\cal V}_{\delta} + {\cal H}_{\text{cc}}  \\
 & {\cal H}_{0} = \frac{2\pi v}{N + 1} \left[ : J^A_{\text{R}} J^A_{\text{R}}: + : J^A_{\text{L}} J^A_{ \text{L}}:  \right]\\
&   {\cal V}_{\delta} =   \bar{\delta} \left(
 \mathrm{e}^{ i (\theta_0 + \frac{\pi}{2} + \frac{\pi }{N})} {\rm Tr} ( g )   + \text{H.c.}  \right)  \\
 & {\cal H}_{\text{cc}} = \lambda_{\text{cc}} J^A_{\text{R}}  J^A_{\text{L}}  ,
\end{split}
\label{eq:contlimitspin}
\end{equation}
where a summation over repeated indices is implied and $\bar{\delta} = \mathcal{C}_N \delta$ ($\mathcal{C}_N >0$); see Supplemental Material for more details (SM)~\cite{supmat}. A similar derivation has been obtained in Ref.~\onlinecite{Lajko-W-M-A-17} in the $N=3$ case. The leading contribution is ${\cal V}_{\delta}$ which is a strongly relevant perturbation with scaling dimension $(N-1)/N$ and ${\cal H}_{\text{cc}}$ is a marginal current-current interaction. A spectral gap is opened for either sign of the modulation $\delta$ with an energy gap, $ \Delta \sim |\bar{\delta}|^{N/(N+1)}/(\ln |\bar{\delta} |) ^{(N-1)/N}$ \cite{supmat}.  The physical nature of the gapped phases strongly depends on the sign of $\delta$.

When $\delta<0$ (i.e., $\bar{\delta}<0$), the minimization of the strongly relevant perturbation ${\cal V}_{\delta}$ in Eq.~(\ref{eq:contlimitspin}) leads to a non-degenerate solution for every $N$:
\begin{equation}
g_{\delta < 0} = \mathrm{e}^{ - i (\theta_0 + \frac{\pi}{2} + \frac{\pi }{N})} I,
\label{gsdelta<0}
\end{equation}
$I$ being the $N \times N$ identity matrix. The solution is an $\SU(N)$ matrix if $\mathrm{e}^{ i N \theta_0}  = - ( - i) ^{N}$, which fixes the phase $\theta_0$ of the non-universal constant that appears in the continuous description of the spin operator (\ref{eq:spinop}). The ground state, described by the solution \eqref{gsdelta<0}, is invariant under the inversion symmetry. The phase is a featureless fully gapped phase which 
is made by a collection of singlet states of $N$ sites (see Fig. \ref{fig:su3}(i)  for $N=3$). 

When $N$ is odd and $\delta > 0$, the minimisation of  the potential  ${\cal V}_{\delta}$ in  Eq. (\ref{eq:contlimitspin}) gives a two-fold degenerate solution:
\begin{equation}
g_{\pm} =  - \mathrm{e}^{ -i (\theta_0 + \frac{\pi}{2} + \frac{\pi }{N}) \pm  i \frac{\pi }{N}} I .
\label{solutiondelta>0oddN}
\end{equation}
Under the inversion symmetry (\ref{eq:contlimitsym}), we have: $g_{\pm}  \xrightarrow{\mathcal{I}} g_{\mp}$. The phase in the odd-$N$ case is thus twofold degenerate and spontaneously breaks the inversion symmetry. From the identification (\ref{eq:contlimitsym}), we observe that the solutions (\ref{solutiondelta>0oddN}) are obtained from the trivial one (\ref{gsdelta<0}) by a simple translation of $(N\pm1)/2$ sites. In an open geometry, the phase with $\delta < 0$ has no edge state whereas the two degenerate phases with $\delta > 0$ enjoy chiral edge states due to the translation of $(N\pm1)/2$ sites with the left edge-state being described  by a Young tableau with a single column and $(N\pm1)/2$ boxes and the right one in the representation with a single column and $(N\mp1)/2$ boxes. These edge states are exchanged by the inversion symmetry and belong to conjugate representations. In the simplest $N=3$ case, the edge states belong to the ${\bf 3}$ and ${\bar {\bf 3}}$ representations of the $\SU(3)$ group and define the $\PSU(3)$ chiral SPT phase which was found in two-leg spin ladders with unequal spins or other 1D $\SU(3)$ spin models in the adjoint representation of the $\SU(3)$ group \cite{Rachel-S-S-T-G-10,Morimoto-U-M-F-14,Ueda-M-M-18,Roy-Q-18,Fromholz-C-L-P-T-19,Capponi-F-L-T-20}. In the general odd-$N$ case, the twofold-degenerate ground state for $\delta >0$ corresponds to the chiral $\frac{(N\pm 1)}{2}$- SPT phases with edge states with dimension: $d =  \frac{N !}{\left[(N -1)/2\right] ! \left[(N +1)/2\right] !}$. The inversion symmetry is spontaneously broken in the chiral SPT phase and remains unbroken in the trivial phase. Despite this $\Z_2$ inversion broken symmetry, the phase transition at $\delta=0$ between the trivial and chiral SPT phases is not an Ising transition with central charge $c=1/2$ but belongs to the $\SU(N)_1$ universality class~\cite{Affleck-88} with a central charge $c=N-1$. The transition satisfies the bound $c\ge \log_2 d$ conjectured in Ref. \onlinecite{Verresen-M-P-17} corresponding to a 1D phase transition driven by a delocalized egde mode with dimension $d$.

\para{Strong coupling}%
In order to restrict ourselves to antiferromagnetic interactions (i.e. positive coupling constants), one has to limit $\delta< \delta_{\mathrm{max}} = 1/\cos(\pi/N)$. Close to this maximal value, the model maps onto an effective $\SU(N)$ chain in the $N-\bar{N}$ representation, which has been extensively studied and is expected to be spontaneously dimerized~\cite{Affleck-SUN-90}.  
For any $N$, Affleck has shown that the $N-\bar{N}$ chain is equivalent to an $N^2$ Potts model~\cite{Affleck-SUN-90} and hence spontaneously dimerized with a finite gap for $N>2$, generalizing the $N=3$ result~\cite{Barber-B-89}. Moreover, by mapping the Potts model onto an $S=1/2$ XXZ chain, it is possible to get the exact spectrum since this model is integrable. Using these mappings, it was possible to obtain exact results for $N=3$, namely a gap about $\Delta\approx0.173$ and a quite large correlation length $\xi \simeq 21$~\cite{Klumper-89,Klumper-90}. For completeness, we reproduce the main formulas of Ref.~\onlinecite{Klumper-90} and also provide numerical values for other $N$ values relevant to our work in SM~\cite{supmat}. Quite remarkably, our numerical results for the correlation lengths obtained for $N=3$ and $N=5$ are in perfect agreement with these predictions.

\para{Numerical simulations}%
%
Let us now confirm the field theory predictions by numerical simulations of the spin chain model [Eq.~\eqref{eq:hamhidaNodd}] using the matrix product state (MPS) toolbox \cite{Cirac2021}. We use infinite MPS \cite{Vanderstraeten2019} with an explicit encoding of the $\SU(N)$ symmetry, providing the most natural and efficient parametrization of the ground state directly in the thermodynamic limit. The MPS approximation is optimized by a combination of the iDMRG and VUMPS algorithms \cite{McCulloch2008, ZaunerStauber2018, Vanderstraeten2019}, and we use software packages \cite{TensorTrack2024,TensorKit2024, MPSKit2023} with generic non-abelian symmetry support. The accuracy of these MPS ground-state approximations is systematically controlled by the bond dimension or truncation threshold. In order to study the critical behaviour around $\delta=0$, we employ accurate techniques \cite{Rams2018, Vanhecke2019} to extrapolate our results to infinite bond dimension. We refer to SM~\cite{supmat} for the details on the MPS simulation techniques. 

First, we discuss some general features for all our $\SU(N)$ simulations. As shown in SM~\cite{supmat}, the manifold of infinite $\SU(N)$-invariant MPS is divided into different classes, where a variational optimization can determine which class is realized for a given value of $\delta$. For all $\delta < 0$, we find that the ground state belongs to class $\class{0}$, which is consistent with a trivial state adiabatically connected to a product state made of $N$-site singlets and the predicted result of Eq.~\eqref{gsdelta<0}. On the other hand, for $\delta > 0$ we find that MPS in the class corresponding to a chiral $\frac{N\pm1}{2}$-SPT phase for odd $N$, in agreement with Eq.~\eqref{solutiondelta>0oddN}.

In Fig.~\ref{fig:xi} we show the correlation length as a function of $|\delta|$ for the case of $N=3$, obtained after extrapolation using data up to a truncation threshold of $\tau=10^{-8}$. From the knowledge of the ground state's class, this is compatible with a trivial phase for $\delta<0$ and a $\frac{N\pm1}{2}$-SPT phase for $\delta>0$. In the scaling region around $\delta=0$, we fit a power law to our numerical estimates, and find exponents that are reasonably close to the expected exponent $3/4$. We attribute the discrepancies to the logarithmic corrections to the power-law behaviour. 

\begin{figure}
    \centering
    \includegraphics[width=0.999\columnwidth]{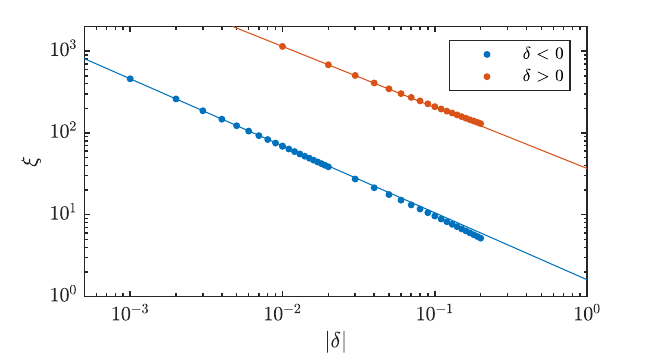}
    \caption{Correlation length vs $|\delta|$ in the SU(3) case for positive and negative $\delta$ (dots), fitted with a power law with exponents $0.824$ and $0.737$ respectively (solid line). The data were obtained from infinite MPS, using proper extrapolation schemes.}
    \label{fig:xi}
\end{figure}

For $\delta>0$ we can measure the bond energies in the unit cell that signal the spontaneous inversion symmetry breaking,  adiabatically connected to the strong coupling limit. We define the order parameter as the difference of the average energy on the  equivalents bonds. For $N=3$, there is a unique order parameter while for $N=5$, we can measure the difference on two pairs of equivalent bonds. As shown in Fig.~\ref{fig:op_sun}, its behavior for $N=3$ is compatible with a power-law with an exponent $0.658$, roughly agreeing with the expected result $1/2$, the difference again probably due to logarithmic corrections.

The numerical simulations become increasingly more demanding for larger $N$, especially in the scaling region around the quantum critical point at $\delta=0$. In SM \cite{supmat} we show additional data for $N=5$, but these are less conclusive than the ones for $N=3$ in Figs.~\ref{fig:xi} and \ref{fig:op_sun}.

\begin{figure}
    \centering
    \includegraphics[width=0.999\columnwidth]{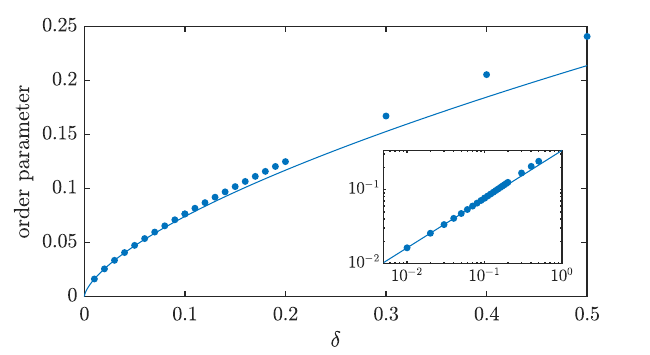}
    \caption{Inversion symmetry order parameter in the SU(3) case for $\delta>0$, evaluated with infinite MPS with a truncation threshold $\tau=10^{-8}$ (dots), fitted with a power law exponent $0.658$ (solid line).}
    \label{fig:op_sun}
\end{figure}

\para{Excitation spectrum}%
As is often the case in 1D spin chains, the symmetry breaking pattern in the ground state determines the nature of the elementary excitations \cite{Su1981, Shastry1981, Faddeev1981}. Here we investigate the spectrum of the $\SU(3)$ chain for $0<\delta<\delta_{\mathrm{max}}=2$, where we have found a chiral SPT phase with two ground states that spontaneously break the inversion symmetry. For large $\delta$, the two symmetry breaking patterns are pictured in Fig.~\ref{fig:su3}: the two spins on the strong bond are always tightly bound into an effective $\threeb$ state, which then dimerizes with the spin to its left or right. One can consider two types of defects in these ground state configurations (Fig.~\ref{fig:su3}), i.e. domain walls \cite{Vanderstraeten2020} carrying either an effective $\three$ or $\threeb$ charge.

The dispersion of these fractional spinon excitations can be calculated numerically using the MPS excitation ansatz, a variational approach for capturing excitations on top of an MPS ground state directly in the thermodynamic limit \cite{Haegeman2012, Vanderstraeten2020}. Then, using the single-spinon dispersion relation, we can simply compute the lower and upper edges of the $s$-$\bar{s}$ continuum. Next, we determine to what extent the full excitation spectrum can be reconstructed from these elementary spinon states. To that effect, we compute the spectral function 
\begin{equation}
    S(q,\omega) = \int_{-\infty}^{+\infty} dt \mathrm{e}^{i\omega t} \bra{\Psi_0} \mathrm{e}^{-iHt} S_{-q}^A \mathrm{e}^{iHt} S_q^A \ket{\Psi_0},
\end{equation}
where $S_q^A$ is the momentum-space spin operator. This spectral function therefore probes the excitations in the $\eight$ (adjoint) sector. We can compute $S(q,\omega)$ by acting with $S_j^\alpha$ on an MPS ground state, performing variational real-time evolution \cite{Zaletel2015, VanDamme2023}, and transforming the correlation function to momentum and frequency space. Results are shown in the left panel of Fig.~\ref{fig:spinons}, as well as the lower and upper edges of the $s$-$\bar{s}$ continuum. We observe that all the dominant features of the spectral function are nicely contained within the continuum, confirming that the spinons are the only fundamental excitations in this system.

As a final application we perturb the modulated $\SU(3)$ spin chain slightly by an extra phase factor $\phi$,
\begin{equation} \label{eq:ham_phi}
    H = J \sum_i \left( 1 + \delta \cos\left(\frac{2\pi i}{N} + \phi \right) \right)S^A_i S^A_{i+1}.
\end{equation}
Such perturbation breaks the inversion symmetry \emph{explicitly}, hence lifting the degeneracy of the two ground states. As a result, the spinons can no longer exist as independent particles, but are confined into bound states. This effect is well-known for (deformed) $\SU(2)$ Heisenberg or Ising chains \cite{Shiba1980, Affleck1998} and has been observed in inelastic neutron scattering experiments \cite{Coldea2010, Lake2010, Bera2017} or after a real-time quench in a quantum simulator \cite{Kormos2017}. In $S(q,\omega)$, this results in the continuum being replaced by isolated lines, which is shown explicitly in the right panel of Fig.~\ref{fig:spinons}. The enthusiastic reader can interpret our current $\SU(3)$ example as a 1D spin chain analog of the confinement of quarks into mesonic bound states.

\begin{figure*}
    \centering
    \includegraphics[width=0.8\textwidth]{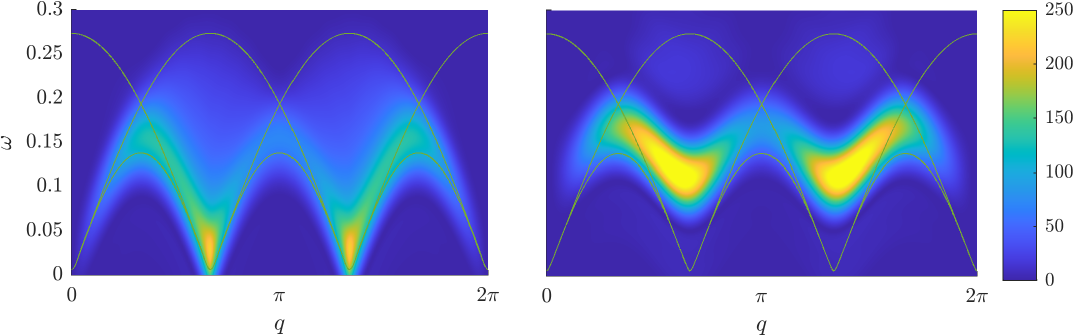}
    \caption{The spectral function $S(q,\omega)$ of the $\SU(3)$ chain with $\delta=1.7$ and $\phi=0$ (left) and $\phi=\pi/100$ (right), with the lower and upper edges of the two-spinon continuum in green.}
    \label{fig:spinons}
\end{figure*}

\para{Summary and experimental realization}%
In this work, we have have shown that the modulated $\SU(N)$ spin chain exhibits a non-Landau quantum phase transition, both from a low-energy field theory approach and from numerical MPS simulations. For odd $N$, the transition is between a trivial phase and a chiral-SPT phase with twofold degenerate ground state. In the latter, the excitations are deconfined spinon excitations with fractional $\SU(N)$ charge, which can be confined by adding a small phase in the modulation of the coupling strengths.

Given these exotic phenomena, it would be very interesting to engineer this system in a quantum simulation experiment. There are several ways to realize an $\SU(N)$-symmetric spin model using alkaline-earth ultracold atoms such as Yb or Sr \cite{Cazalilla-R-14}. For simplicity, we consider only the $N=3$ case, in which we can start from an optical lattice made of a set of three laser beams with the wavelengths $2\Lambda$, $\Lambda$, and $\Lambda/3$: 
\begin{multline}
V_{\text{lat}}(x)  
 = 
V_{1} \cos^{2} \left( \frac{2\pi}{2\Lambda}x\right) \\ + V_{2} \cos^{2} \left( \frac{2\pi}{\Lambda}x\right)
+ V_{3} \cos^{2} \left( \frac{2\pi}{2\Lambda/3}x\right)  \; .
\end{multline}
For appropriate choices of the beam strengths $V_{1,2,3}$, we obtain a lattice with three minima within a period $\Lambda$ separated by large potential barriers. Now we load fermionic atoms into the lattice and carry out the second-order perturbation in the hopping to arrive at an $\SU(3)$ Heisenberg chain with modulated interactions. We can realize the model \eqref{eq:hamhidaNodd} with $\delta <0$ by choosing large enough $V_{2} \,(>0)$ and setting $V_{1} =r V_{2}$ ($r > 1$), and $V_{3} = V_{2}/[4(r - 1)]$. The last condition for $J_{3}$ ({\em triple-well condition}) is necessary for the three minima to have equal depths. On the other hand, to simulate the model with $\delta >0$, we use large enough $V_{2} \,(< 0)$ and $V_{1} = r V_{2}\, (<0)$, and tune the third beam as: $V_{3} = |V_{2}| /[4(1- r)]$ ($0 < r <1$). We can effectively flip the signs of $V_{1}$ and $V_{2}$ by introducing phase shifts in the corresponding beams.

\para{Acknowledgments}%
We would like to thank E. Orignac  and D. Papoular for useful discussions. S. C. acknowledges the use of HPC resources from CALMIP (Grant No. 2023-P0677) and GENCI (Project No. A0130500225). The authors would like to thank CNRS and acknowledge the financial support (grant CNRS IRP EXQMS). L. D. is supported by Research Foundation Flanders (FWO) via grant 3G0E1820. L. V. is supported by the Fonds de la Recherche Scientifique de Belgique (F.R.S.-FNRS). 

%


\pagebreak
\clearpage

\begin{widetext}
\begin{center}
\textbf{\large Supplemental Material for ``Non-Landau quantum phase transition \\ in modulated SU(\textit{N}) Heisenberg spin chains''}
\end{center}
\end{widetext}

\author{S. Capponi} 
\affiliation{Laboratoire de Physique Th\'eorique, Universit\'e de Toulouse, CNRS, UPS, France.}

\author{L. Devos}
\affiliation{Department of Physics and Astronomy, Ghent University, Belgium}

\author{P. Lecheminant} 
\affiliation{Laboratoire de Physique Th\'eorique et Mod\'elisation, CNRS, CY Cergy Paris Universit\'e, 95302 Cergy-Pontoise Cedex, France.}

\author{K. Totsuka} 
\affiliation{Center for Gravitational Physics and Quantum Information, 
Yukawa Institute for Theoretical Physics, Kyoto University, Kyoto 606-8502, Japan.}

\author{L. Vanderstraeten}
\affiliation{Center for Nonlinear Phenomena and Complex Systems, Universit\'e Libre de Bruxelles, Belgium}

\maketitle

\section{Continuum description}\label{app:Contlimit}

Here we derive the continuum limit of the $\SU(N)$ modulated spin chain (\ref{eq:hamhidaNodd}) and  of the order parameter which describes the spontaneous breaking of the inversion symmetry in the $\delta >0$ phase. For completeness, we discuss also the even-$N$ case when $\delta >0$ phase with the emergence of a non-degenerate 
$N/2$-SPT phase.

\subsection{Continuum limit of the lattice Hamiltonian}

We first rewrite the Hamiltonian (\ref{eq:hamhidaNodd}) as 
\begin{equation}
 \begin{split}
 H_{\delta}  &=     H_{SU(N)} +  H_{\delta} \\
 & = J \sum_{i=1}^{NL} S^{A}_{i+1} S^{A}_{i}   + 
 J \delta \sum_{i=1}^{NL} \cos \left(\frac{2 \pi i}{N} \right) \;  S^{A}_{i+1} S^{A}_{i} ,
 \end{split}
\label{eq:hamdeltacont}
\end{equation}
where a summation over repeated SU($N$) indices ($A$)  is implied. The decomposition (\ref{eq:hamdeltacont}) enables us to perform a continuum limit of ${\cal V}_{\delta}$ in the vicinity of the SU($N$)$_1$ quantum critical point obtained when $\delta =0$. In the low-energy limit, the lattice spin operators are described by \cite{Affleck-NP86,Affleck-88,James-K-L-R-T-18} 
\begin{equation}
S^{A}_{n}/a_0 \simeq  J^{A}_{\text{L}} +  J^{A}_{\text{R}} 
+  i  \lambda \, \be^{\frac{ i 2 \pi }{Na_0}x} \; {\rm Tr} ( g (x)  T^A) + \text{H.c.},
\label{eq:spinopapp}
\end{equation}
where $x=n a_0$ ($a_0$ being the lattice spacing), $J^{A}_{\text{R,L}}$ are the chiral SU($N$)$_1$ currents, $g$ is the $\SU(N)_{1}$ primary field with the scaling dimension $(N-1)/N$ and $\lambda = Ce^{ i \theta_0}$ ($C >0$) is a non-universal complex constant which stems from the averaging of the underlying charge degrees of freedom which are frozen in the insulating phase of the SU($N$) Heisenberg spin chain. 

\begin{widetext}
Using the identification \eqref{eq:spinopapp} and dropping the marginal current-current contribution as well as the oscillatory terms, 
one obtains the following density $\mathcal{V}_{\delta}$ corresponding to $H_{\delta}$:
\begin{equation}
 \begin{split}
 {\cal V}_{\delta} &=   \frac{J  \delta a_0}{2} \mathrm{e}^{-i \frac{2 \pi n}{N}} \left[
 J^{A}_{R} + J^{A}_{L} + i  C \mathrm{e}^{i \frac{2 \pi n}{N} } \mathrm{e}^{i \frac{2 \pi }{N} }  \, \mathrm{e}^{i \theta_0}
  \, {\rm Tr} ( g  T^A)  - i C \mathrm{e}^{-i \frac{2 \pi n}{N} } \mathrm{e}^{-i \frac{2 \pi}{N} }  \, e^{-i \theta_0}
  \, {\rm Tr} ( g^{\dagger}  T^A)\right]  (x+a_0) \\
  & \qquad \times  \left[
 J^{A}_{R} + J^{A}_{L} + i C e^{i \frac{2 \pi n}{N} } \, \mathrm{e}^{i \theta_0}
  \, {\rm Tr} ( g  T^A)   - i C e^{-i \frac{2 \pi n}{N} }  \, e^{-i \theta_0}
  \, {\rm Tr} ( g^{\dagger}  T^A)\right] (x) + \text{H.c.} \\
 & \simeq \frac{ i J C  a_0 e^{i \theta_0}}{2} \delta \left[  \left(J^{A}_{R} + J^{A}_{L} \right) (x+a_0) 
   \; {\rm Tr} ( g  T^A) (x)  
   + \mathrm{e}^{i \frac{2 \pi }{N} }   \; {\rm Tr} ( g  T^A) (x+a_0)  \left(J^{A}_{R} + J^{A}_{L} \right) (x)\right] 
   + \text{H.c.}  \; .
 \end{split}
\label{eq:vdeltacont}
\end{equation}
\end{widetext}
To simplify the above, we need the following operator product expansions (OPE) \cite{DiFrancesco-M-S-book}: 
\begin{equation}
\begin{split}
& J_{L}^A\left(z\right)  {\rm Tr} ( g T^{A} ) (0,0) \sim - \frac{N^2-1}{4 \pi N z} \; 
{\rm Tr} ( g)  (0,0)  \\
& J_{R}^A\left(\bar z\right)  {\rm Tr} ( g T^{A} ) (0,0) \sim \frac{N^2-1}{4 \pi N \bar z} \; 
{\rm Tr} ( g)  (0,0) \\
& {\rm Tr} ( g T^{A} ) \left(z,\bar z\right)    J_{L}^A (0) \sim  \frac{N^2-1}{4 \pi N z} \; 
{\rm Tr} ( g)  (0,0)  \\
& {\rm Tr} ( g T^{A} ) \left(z,\bar z\right)    J_{R}^A (0)  \sim - \frac{N^2-1}{4 \pi N \bar z} \; 
{\rm Tr} ( g)  (0,0)  \\
& J_{L}^A\left(z\right)  {\rm Tr} ( g^{\dagger} T^{A} ) (0,0) \sim  \frac{N^2-1}{4 \pi N z} \; 
{\rm Tr} ( g^{\dagger})  (0,0)  \\
& J_{R}^A\left(\bar z\right)  {\rm Tr} ( g^{\dagger} T^{A} ) (0,0) \sim  - \frac{N^2-1}{4 \pi N \bar z} \; 
{\rm Tr} ( g^{\dagger})  (0,0)  \\
& {\rm Tr} ( g^{\dagger} T^{A} ) \left(z,\bar z\right)    J_{L}^A (0) \sim - \frac{N^2-1}{4 \pi N z} \; 
{\rm Tr} ( g^{\dagger})  (0,0)  \\
& {\rm Tr} ( g^{\dagger} T^{A} ) \left(z,\bar z\right)    J_{R}^A (0) \sim  \frac{N^2-1}{4 \pi N \bar z} \; 
{\rm Tr} ( g^{\dagger})  (0,0) \;  ,
\end{split}
\label{definingtraceOPEs}
\end{equation}
with $z= v\tau + ix, \bar z = v\tau - ix$ ($v$ being the spin velocity).

The leading contribution of Eq. (\ref{eq:vdeltacont}) is then 
\begin{equation}
 \begin{split}
{\cal V}_{\delta} &\simeq  \frac{J (N^2-1) C \mathrm{e}^{i \theta_0}}{4 \pi N} \delta 
\left( \mathrm{e}^{i \frac{2 \pi }{N}} - 1\right) 
 \; {\rm Tr} ( g ) + \text{ H.c. } \\
 & = \bar{\delta} \left(
 \mathrm{e}^{ i \theta_0 + i \frac{\pi}{2} + i \frac{\pi }{N}} {\rm Tr} ( g )   + \text{H.c.}  \right) ,
 \end{split}
\label{eq:vdeltacontfin}
\end{equation}
where we have introduced $\bar{\delta}= \mathcal{C}_N \delta \, (\ll 1)$ with the positive constant defined by:
\begin{equation}
\mathcal{C}_N = \frac{J (N^2-1) C \sin\left( \frac{\pi }{N}\right)}{2 \pi N} \, (>0) \; .
\end{equation}

A spectral gap $\Delta$ is generated by the relevant perturbation with scaling dimension $1-1/N$ when $\delta \ne 0$. The correlation length is finite and scales as function of the coupling constant $\delta$:
\begin{equation}
\xi \sim  |\bar{\delta}|^{-N/(N+1)} ,
\label{estimatecorrelength}
\end{equation}
so that we have a gap $\Delta \sim |\bar{\delta}|^{N/(N+1)}$. Logarithmic corrections are also expected in the latter expression since the subleading term in the Hamiltonian is marginal \cite{Affleck-G-S-Z-89}. These corrections have been investigated in SU($N$) spin chain in Ref.
\onlinecite{Majumdar-M-02}. We get in our case:
\begin{equation}
 \Delta \sim |\bar{\delta}|^{N/(N+1)}(\ln | \bar{\delta} |) ^{-(N-1)/N} .
\label{estimategap}
\end{equation} 

\subsection{Continuum limit of the order parameter}

We now consider the order parameter used in the numerical simulations to detect the inversion symmetry breaking. An order parameter which measures the inversion-symmetry breaking in the chiral SPT phases can be defined as follows:
\begin{equation}
 {\cal O}_n   =     S^{A}_{Nn+1} S^{A}_{Nn+2}   - S^{A}_{Nn+N-1} S^{A}_{Nn+N} .
\label{eq:orderparainvbreak}
\end{equation}
It corresponds to the order parameter used in the numerical simulations being the average energy on equivalent bonds.
Its continuous description can be obtained by means of the identification (\ref{eq:spinopapp}) and the use of the OPEs (\ref{definingtraceOPEs}). The derivation is similar to the continuum limit of the lattice Hamiltonian and we get after some cumbersome calculations:
\begin{equation}
  {\cal O} \simeq    \frac{2 C (N^2-1)}{\pi N a_0} \sin \Bigl(\frac{2\pi}{N} \Bigr)
  \sin \Bigl(\frac{\pi}{N}\Bigr) \mathrm{e}^{ i (\theta_0  + \frac{\pi }{N})} {\rm Tr} ( g )   + \text{H.c.} 
\label{eq: orderparacont}
\end{equation}
In the $\delta<0$ ($\bar{\delta} <0 $) phase, we have $g = \mathrm{e}^{ - i (\theta_0 + \frac{\pi}{2} + \frac{\pi }{N})} I$ (see Eq. (\ref{gsdelta<0})) and $\langle {\cal O} \rangle = 0$. In contrast, the ground state is two-fold degenerate in the $\delta > 0 $ phase  with
$g_{\pm} =  - \mathrm{e}^{ -i (\theta_0 + \frac{\pi}{2} + \frac{\pi }{N}) \pm  i \frac{\pi }{N}} I$ (see Eq. (\ref{solutiondelta>0oddN})). The order parameter (\ref{eq: orderparacont}) condenses in these configurations:
\begin{equation}
  \langle {\cal O} \rangle_{\pm} \simeq  \mp  \frac{ 4 C (N^2-1)}{\pi a_0} \sin \Bigl(\frac{2\pi}{N} \Bigr)
  \sin^2 \Bigl(\frac{\pi}{N}\Bigr) ,
\label{eq: orderparacondens}
\end{equation}
which marks the spontaneous-breaking of the inversion symmetry in the
 $\delta > 0 $ phase. 

If we note that the ground state energy density $e_{0}$ is scaled by 
the correlation length $\xi$ as $e_{0} \sim \xi^{-2}$, 
we can deduce the scaling form of the order parameter as a function of the coupling constant $\bar{\delta} >0$.  
Using Eq.~\eqref{estimatecorrelength}, we obtain:
\begin{equation}
   \langle {\cal O} \rangle \sim  \frac{\partial e_{0}}{\partial \bar{\delta}}     
\sim \bar{\delta}^{\frac{N-1}{N+1}} \; . 
\label{eq:orderparainvbreak2}
\end{equation}
As in Eq.~\eqref{estimategap}, the expression acquires logarithmic correction due to the marginal operator: 
\begin{equation}
\langle {\cal O} \rangle \sim 
\bar{\delta}^{\frac{N-1}{N+1}} / | \log \bar{\delta} |^{2-\frac{2}{N}} \; , 
\end{equation}
when keeping the leading term.

\subsection{Even-$N$ case and $\delta > 0$}

For completeness, we discuss here the even-$N$ case when $\delta > 0$.
When $N$ is even, one can exploit the symmetry of the perturbation (\ref{eq:contlimitspin}) $\delta \rightarrow - \delta$ 
($\bar{\delta} \rightarrow - \bar{\delta}$) and $g  \rightarrow - g$, which is an SU($N$) matrix for even $N$. The phase for $\delta > 0$ is thus non-degenerate as for $\delta < 0$ and it is described in the low-energy approach by the ground-state configuration:
\begin{equation}
g^{N=2p}_{\delta > 0} = - 
\mathrm{e}^{ - i (\theta_0 + \frac{\pi}{2} + \frac{\pi }{N})} I,
\label{gsdelta>0Neven}
\end{equation}
which is invariant under the inversion symmetry (\ref{eq:contlimitsym}).  We note that this solution is obtained from Eq. (\ref{gsdelta<0}) by applying $N/2$ one-step translation symmetry ($T_{Na_0/2}$) which indeed changes the sign of $g$ (see Eq. (\ref{eq:contlimitsym})). Physically, it means that this phase is a collection of singlet states that is obtained from the trivial phase with $\delta < 0$ by shifting the pattern by $T_{Na_0/2}$. In a chain with open boundary conditions, the trivial phase has no edge state. Shifting the singlet pattern by $T_{Na_0/2}$ leads to phase with $N/2$ spins which are not in an SU($N$) singlet at each end of the chain. The non-degenerate phase for $\delta > 0$ is thus a SPT phase with edge states in the self-conjugate antisymmetric representation of SU($N$) which is described by a Young tableau with a single column and $N/2$ boxes. It corresponds to a $N/2$-SPT phase protected by the PSU($N$) symmetry, first identified in Refs. \onlinecite{Nonne2013,Bois2015}. In the simplest $N=2$ case, the phase is adiabatically connected to the spin-1 Haldane phase \cite{Haldane-Novel-lec-17,Hida-92a}. The dimension of the 
Hilbert space of the edge state is $d= N !/\left[(\frac{N}{2})!\right]^2$ and the central charge of the quantum phase transition satisfies the bound $c=N-1\ge \log_2 d$ predicted in Ref.~\onlinecite{Verresen-M-P-17} which stems from the delocalization of this edge state.

\section{MPS with SU(N) symmetry}
\label{app:mps}

\newcommand{\diagram}[1]{\vcenter{\hbox{\includegraphics[scale=0.8,page=#1]{figures/SPT_MPSTensor.pdf}}}}

In order to efficiently compute numerical results, we approximate the ground states in the thermodynamic limit using uniform MPS, possibly with a non-trivial unit cell:
\begin{multline}
    \ket{\Psi(\{A_i\})} = \\ \diagram{5}
\end{multline}
As the Hamiltonian is symmetric under global transformations according to an $\SU(N)$ symmetry, we propose a ground state ansatz that is manifestly symmetric. This constraint translates directly to a symmetry constraint on the local MPS tensors \cite{Cirac2021}: The structure of the virtual spaces on the MPS tensors will also transform according to an $\SU(N)$ representation, which decomposes into a number of different irreducible representations (irreps). Each irrep occurs a number of times. Therefore, in order to find faithful MPS representations for the ground state, we need to determine the size of the unit cell, which $\SU(N)$ irreps occur on a given virtual bond, and the degeneracy of each irrep.

Importantly, the irreps of $\SU(N)$ can be partitioned into $N$ different classes, depending on the total number of boxes that appear in the Young tableaux, modulo $N$. If we denote these classes of irreps as $\lfloor0\rfloor, \lfloor1\rfloor \ldots \lfloor N-1 \rfloor$, we can see that the fusion rules are graded by this partition, or in other words $\lfloor a \rfloor \otimes \lfloor b \rfloor \rightarrow \lfloor a + b \mod{N} \rfloor $.
This has important consequences for the structure of the MPS. As the local Hilbert spaces of the considered models transform according to the fundamental irrep, which belongs to the class $\class{1}$, the total Hilbert space of a symmetric MPS splits into a global superposition of $N$ distinct states:
\begin{equation}
\label{eq:sunsplit}
\begin{split}
    \ket{\Psi(A)} = & \;\diagram{3}\; \\ + & \;\diagram{4}\; \\ + & \; \ldots \;.
\end{split}
\end{equation}
These states do not mix under the application of any symmetric local operator. When the full $\SU(N)$ symmetry is imposed upon the local tensors, we should thus impose the MPS to belong to a single class of these states by restricting the irreps that appear on the virtual level, and simulate separately these $N$ possible classes. We will denote these classes by the representative class on the virtual bond to the left of the first site in the unit cell $\class{a}$. It is interesting to note here that the action of the inversion symmetry exchanges classes $\class{a}$ and $\class{N-a}$.

In order to diagnose the class of the ground state, we can thus optimize and compare each of these $N$ different classes. For equivalent bond dimensions, we can expect that the obtained energy will be lowest if we impose the correct class. However, we can observe an interesting result of the variational minimization: the states are all able to mimic the correct ground state class, by using a specific structure of the MPS tensors. Denoting the ground state tensors from class $\class{a}$ as $A$, it is possible to construct new tensors $\tilde{A}$ belonging to any other class $\class{a+b}$ by taking the tensor product with a purely virtual identity operator $I_{\class{b}}$, where the fusion of the different virtual spaces changes the \emph{effective class}.
\begin{equation}
    \; \diagram{1} \;\coloneqq \;\diagram{2} \;
\end{equation}
Since this extra leg is purely virtual, all expectation values for the MPS with the new tensor $\tilde{A}$ will be exactly the same and, specifically, will have the same variational energy. This does come at a cost, as the effective bond dimension is multiplied with a factor that is equal to the dimension of that identity operator, which is the dimension of the smallest irrep of that class $d_b$. Similarly, we find a signature of this virtual identity operator both in the degeneracies of the entanglement spectrum (the Schmidt values will also appear $d_b$ times) and in the transfer matrix spectrum (there will be a degenerate eigenvalue of magnitude $1$, and the corresponding eigenvectors will have either a trivial charge or a charge $\class{b}$). Because of these degeneracies, we can detect this behaviour very clearly, as the entanglement entropy increases by $\log{d_b}$ as compared to the ground state entanglement entropy, and the MPS will become non-injective. The exact value of this increase that we find in our numerics is also a clear testament of the correctness of this intuitive picture.

Using these manifestly $\SU(N)$-symmetric MPS, we obtain the ground states through a combination of IDMRG and VUMPS \cite{McCulloch2008, ZaunerStauber2018}. The bond dimension within each virtual irrep is determined by a two-site update scheme, and truncate the bond dimension up to a certain thruncation threshold $\tau$; the value of $\tau$ therefore determines the bond dimension of the MPS. 

\begin{figure}
    \centering
    \includegraphics[width=0.999\columnwidth]{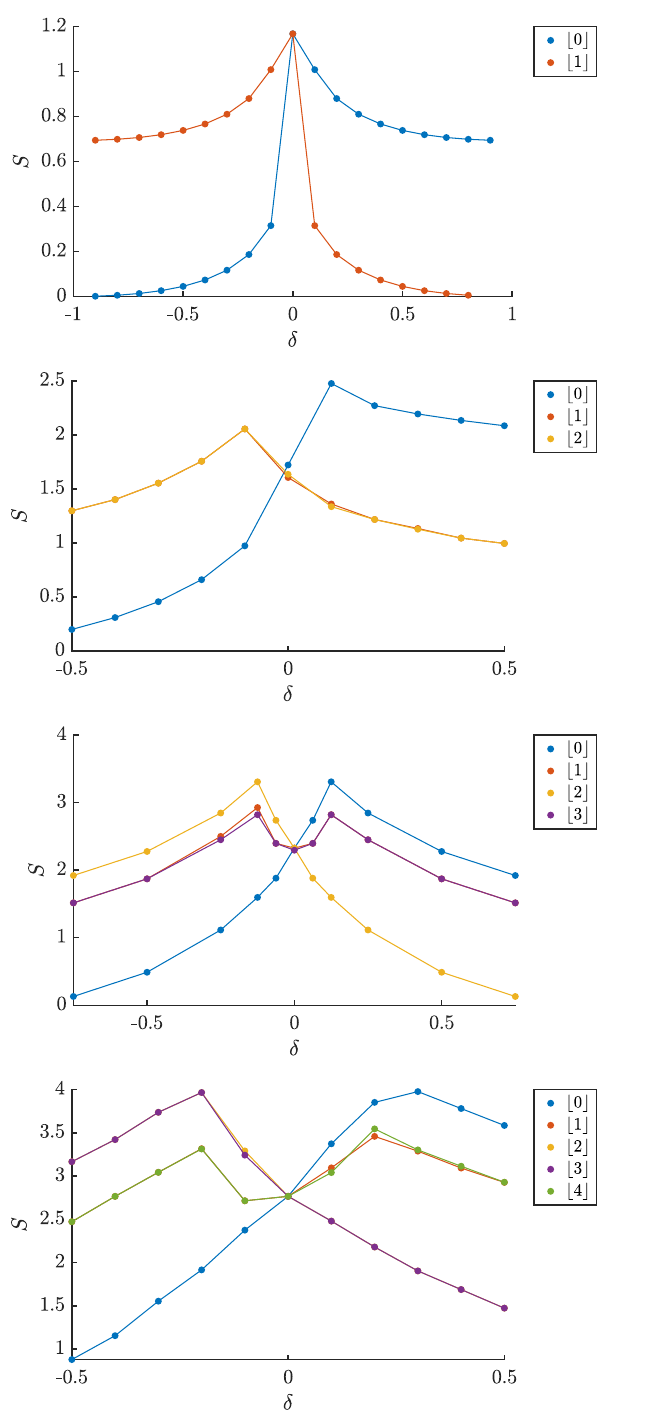}
    \caption{Half-system entanglement entropy vs $\delta$ for the $N=2$, $3$, $4$ and $5$ cases (from top to bottom) using different classes of virtual spaces (see text).}
    \label{fig:check_spt}
\end{figure}

This is done through the open-source libraries \cite{TensorKit2024, MPSKit2023, TensorTrack2024}. The results in Fig.~\ref{fig:check_spt} clearly demonstrate that for $\delta < 0$, the ground state belongs to class $\class{0}$, which is consistent with singlets forming between sites $1$ to $N$, and the predicted result of Eq.~\eqref{gsdelta<0}. On the other hand, for $\delta > 0$ our numerical results show unambiguously that the ground state belongs to the class $\class{N / 2}$ for even $N$, in agreement with Eq.~\eqref{gsdelta>0Neven} and to the 
classes $\class{\frac{N\pm1}{2}}$ for odd $N$, in agreement with Eq.~(\ref{solutiondelta>0oddN}). These results are in line with those obtained from the low-energy approach obtained from the main text.

\section{Extrapolating MPS results}

In the scaling region around the quantum critical point, the MPS results will suffer from artefacts due to finite bond dimension (or, in our simulations, finite truncation threshold $\tau$). Fortunately, we can use strong extrapolation techniques to obtain reliable results, and show the correct scaling behaviour.
\par
\begin{figure}
    \centering
    \includegraphics[width=0.999\columnwidth]{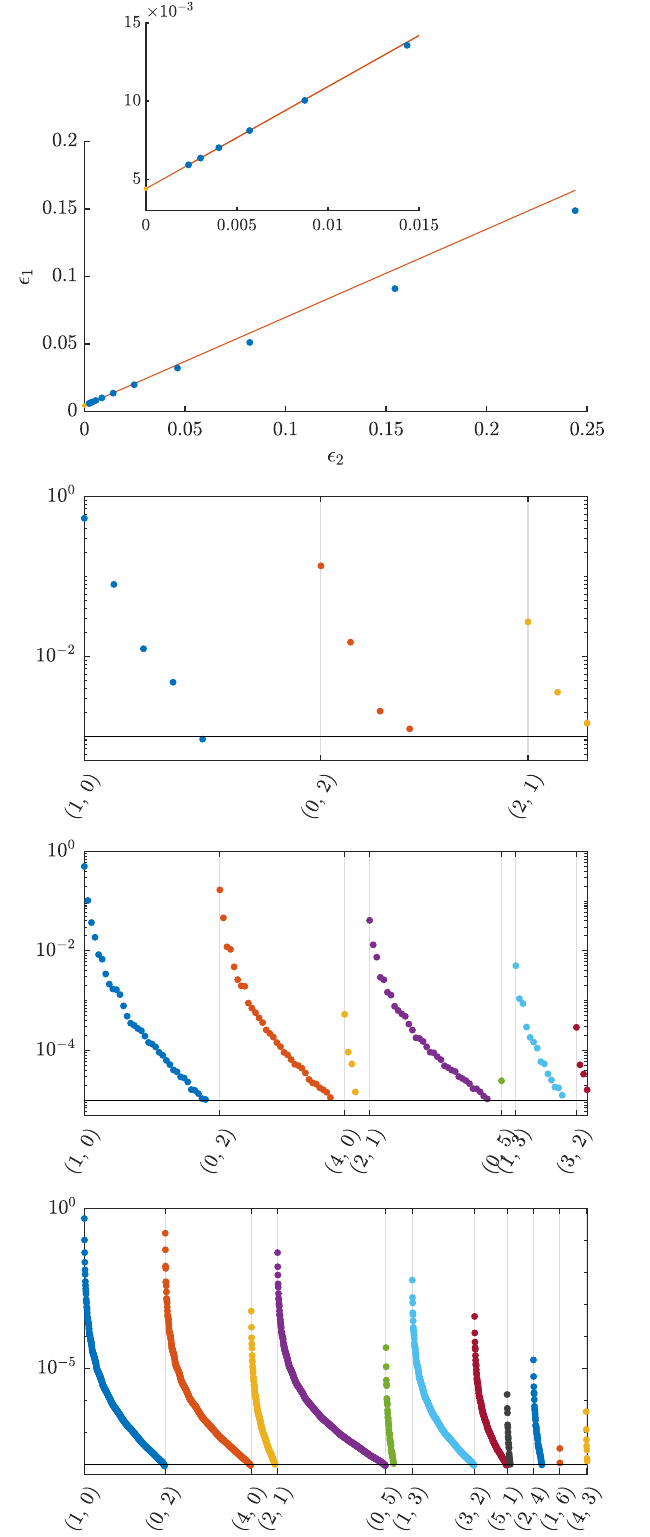}
    \caption{Extrapolation of the correlation length for the $\SU(3)$ chain at $\delta=0.02$. In the top panel, the $(\epsilon_1,\epsilon_2)$ data is shown, along with a fit that allows us to estimate the correlation length. Fitting the last four points, we find $\xi=227.04\pm0.45$. In the bottom three panels, we show the MPS entanglement spectra, labelled by irreps using Dynkin label, for different truncation thresholds $\tau=(10^{-3},10^{-5},10^{-8})$ (the horizontal lines show the thresholds). The largest total bond dimension is around $D\approx4\times10^4$.}
    \label{fig:extrapolation}
\end{figure}

 As an illustration, we take the $\SU(3)$ chain with $\delta=0.02$, very close to the critical point. For every value of the truncation threshold $\tau$, we find an $\SU(3)$-invariant MPS approximation of the ground state with a given bond dimension within each virtual irrep. This distribution of bond dimensions is illustrated for different values of $\tau$ in the bottom panels of Fig. \ref{fig:extrapolation}. We use the spectrum of the MPS transfer matrix \cite{Rams2018, Vanhecke2019} to extrapolate: The leading eigenvalue is always normalized to one, whereas the subleading eigenvalues $\lambda_i$ can be used for extrapolation as follows. We can extract a first gap $\epsilon_1=-\log(|\lambda_1|)$, which is related to the correlation length of the MPS as $\xi/N=1/\epsilon_1$ (where $N$ is the size of the unit cell of the MPS). We can also extract the logarithm of the second gap in the transfer matrix $\epsilon_2=-\log(|\lambda_2/\lambda_1|)$. Gathering $(\epsilon_1,\epsilon_2)_\tau$ values for decreasing values of the truncation threshold $\tau$, we can fit to the form \cite{Rams2018}
\begin{equation} \label{eq:marek}
\epsilon_1(\tau) = \epsilon_{\tau\to0} + a \epsilon_2(\tau),
\end{equation}
from which we can extract an extrapolated estimate for the correlation length. In Fig.~\ref{fig:extrapolation} we show the fitting procedure, as well as three representative MPS entanglement spectra.

\section{SU(2) case as a benchmark}\label{app:su2}
For $N=2$, our model is simply the spin-1/2 Heisenberg chain with explicit dimerization $\delta$ and different boundary conditions depending on its sign. It is well-known that any finite $\delta$ will open a finite spin gap $\Delta$ and conversely a finite correlation length $\xi \sim 1/\Delta \sim |\delta|^{-2/3}$. Due to the presence of a marginal term, there also appears significant logarithmic corrections that could be important in the numerical analysis~\cite{Papenbrock2003,Orignac2004} so that spin gap data could be fitted using
\begin{equation}
    \Delta = \alpha^{1/2}_\mathrm{gap} \frac{\delta^{2/3}}{(\ln \delta_0/\delta)^{1/2}}
\end{equation}
with $\alpha_\mathrm{gap}=19.4$ and $\delta_0=115$.

\begin{figure}
    \centering
    \includegraphics[width=0.999\columnwidth]{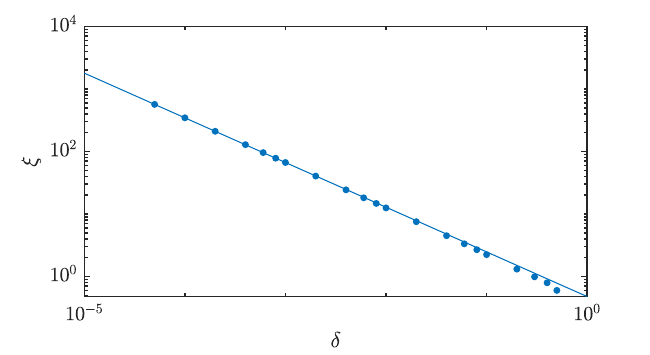}
    \caption{Correlation length vs $\delta$ in the SU(2) case; we find an exponent around 0.71. Data were obtained using extrapolation (see text).}
    \label{fig:xi_su2}
\end{figure}

In our approach, we can directly obtain the correlation length for a given $\delta$ after proper extrapolation~\cite{Rams2018}. Note that it is not expected to scale exactly as the inverse gap. Our data are shown in Fig.~\ref{fig:xi_su2}. They can be nicely fitted using a single power-law as $1/\delta^{0.71}$ but also with the expected analytical exponent $2/3$ when logarithmic corrections are included:
\begin{equation}
    \xi(\delta) \sim 0.24 (\log(2.75/\delta) )^{1/2} /\delta^{2/3} 
\end{equation}

\section{SU(4) case}\label{app:su4}
While our paper focuses on the odd $N$ case where there is a spontaneous inversion symmetry breaking for $\delta>0$, it can be useful to contrast this behaviour with respect to the even $N$ case. In this case, one can change the sign of $\delta$ by a simple translation of $N/2$ sites. This maps a trivial singlet into a non-degenerate $\class{N/2}$ ground state with nontrivial edge states (that transform into the self-conjugate irrep) in full agreement with the low-energy approach.

\begin{figure}
    \centering
    \includegraphics[width=0.999\columnwidth]{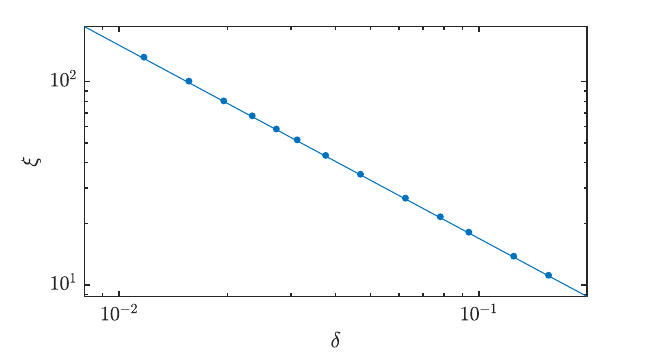}
    \caption{Correlation length vs $\delta$ in the SU(4) case; we find an exponent around 0.95. Data were obtained using extrapolation.}
    \label{fig:xi_su4}
\end{figure}

As discussed for $N=2$ in the previous section, the correlation length is identical for $\pm \delta$ and its behavior for $N=4$ is shown in Fig.~\ref{fig:xi_su4}. Its scaling divergence as $1/\delta^{0.95}$ is in poor agreement with the analytical expectation $N/(N+1)=0.8$, probably due to logarithmic corrections, as we have observed for SU(2) in the previous section.

\section{SU(5) case}\label{app:su5}
\begin{figure}
    \centering
    \includegraphics[width=0.999\columnwidth]{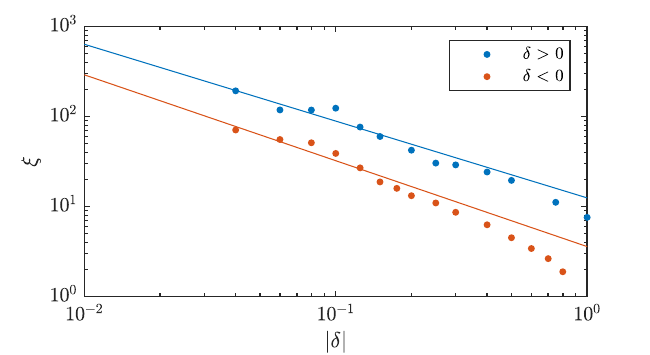}
    \caption{Correlation length vs $|\delta|$ in the SU(5) case for positive and negative $\delta$. We find exponents $0.85$ and $0.95$ respectively. Data were obtained using extrapolation (see text).}
    \label{fig:xi_su5}
\end{figure}

For completeness, we present in this section some data obtained for $N=5$. Due to the growing complexity in the simulations, we could not fully converge the data at small $\delta$ (we had to discard states with truncation thresholds smaller than $10^{-4}$ typically) so that accuracy on the inversion symmetry order parameter is not high. The correlation lengths, obtained after extrapolation (see above), are shown in Fig.~\ref{fig:xi_su5} and are in rough agreement with the expected power-law $|\delta|^{-5/6}$, probably due to logarithmic corrections in the scaling or to our poor convergence.

\begin{figure}
    \centering
    \includegraphics[width=0.999\columnwidth]{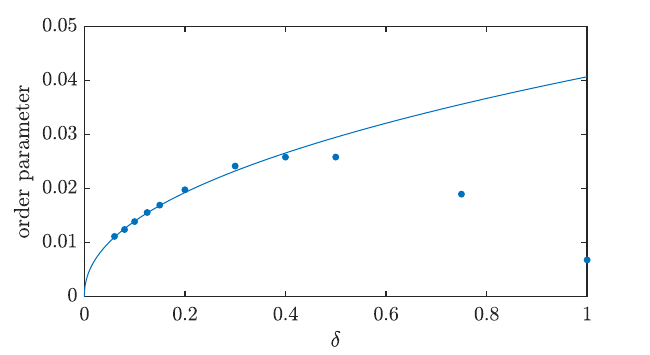}
    \caption{Inversion symmetry order parameter in the  SU(5) case for $\delta>0$, fitted using a power law with exponent $0.658$.}
    \label{fig:op_su5}
\end{figure}

Regarding the inversion symmetry breaking for $\delta>0$, we do measure unambiguously different bond energies on equivalent bonds so that we can plot e.g. the order parameter (\ref{eq:orderparainvbreak}) in Fig.~\ref{fig:op_su5}. This quantity is vanishing when $\delta$ gets close to $\delta_\mathrm{max}$. For small $\delta$, we can fit a power-law with an exponent consistent with the expected one $(N-1)/(N+1)=2/3$.

\section{$N$-$\bar{N}$ chain}\label{app:NNbar}

When $\delta$ is close to its maximal value $\delta_{\mathrm{max}}$ and for odd $N$, the model maps onto an effective SU($N$) chain (with an effective coupling simply given by $\delta_{\mathrm{max}}-\delta$) where each site transforms in the fundamental (respectively conjugate) irrep alternatively, the so-called $N$-$\bar{N}$ chain. In particular, the $N=3$ case can be mapped to a $\SU(2)$ spin-1 chain with a purely negative biquadratic interaction~\cite{Barber-B-89,Sorensen1990}.

As discussed in the main text, for any $N$ this system is known to be spontaneously dimerized and gapped. Following~\cite{Klumper-89,Klumper-90}, let us define
\begin{equation}
    \omega=\frac{(1 + \sqrt{\frac{N + 2}{N - 2}})}{2}\qquad \alpha=\left(\frac{\omega}{\omega-1}\right)^2=(N-2)^2\omega^4
\end{equation}
Then, the gap is given by
\begin{equation}
    \Delta_N = \sqrt{N^2-4} \prod_{n=1}^\infty \left( \frac{1-(1/\alpha)^{n/2}}{1+(1/\alpha)^{n/2}}\right)^2
\end{equation}
For the correlation length, one needs first to solve the equation:
\begin{equation}
      k = \frac{4}{\sqrt\alpha}
 \prod_{n=1}^\infty \left(\frac{1 + (1/\alpha)^{2n}}{1 + (1/\alpha)^{2n - 1}}\right)^4
\end{equation}
related to complete elliptic integral of the first kind~\cite{Baxter_book}, and then the correlation length is given by $\xi_N = -1/\log k$.

For $N=3$, we recover the known values~\cite{Klumper-90}: $\Delta_3\simeq 0.173$ and $\xi_3\simeq 21$. Note that using similar techniques, another reference has found a correlation length twice larger~\cite{Sorensen1990} but subtle issues with the transfer matrix spectrum can lead to an additional factor 2, see e.g.~\cite{Baxter_book}. For $N=5$, we find $\Delta_5\simeq 1.576$ and $\xi_5\simeq 2.9$. As expected for increasing $N$, the correlation length becomes shorter and the gap larger.

\begin{figure}
    \centering
    \includegraphics[width=0.999\columnwidth]{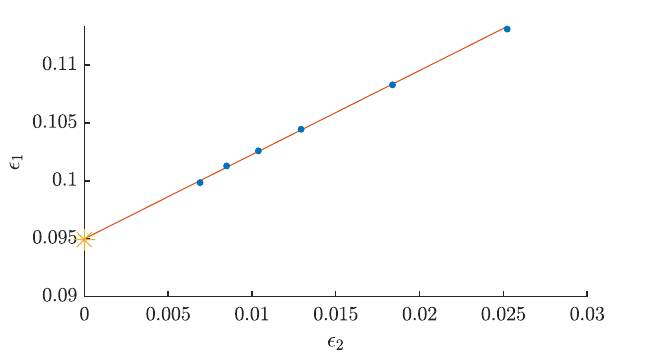}
    \caption{Extrapolation of the normalized inverse correlation length $\epsilon_1=2/\xi$ for the $3-\bar{3}$ SU(3) chain, according to Eq.~\eqref{eq:marek}. The extrapolated value is very close to the exact result (yellow star).}
    \label{fig:xi_33bar}
\end{figure}

\begin{figure}
    \centering
    \includegraphics[width=0.999\columnwidth]{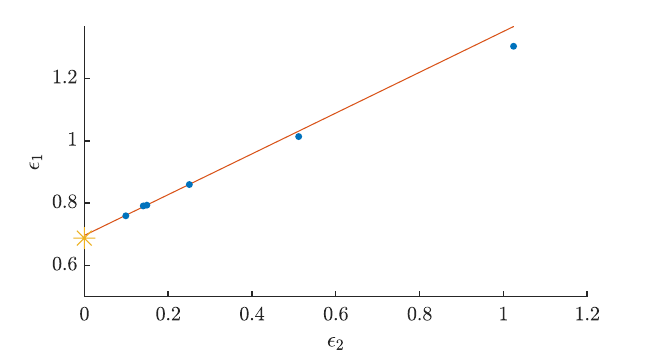}
    \caption{Extrapolation of the normalized inverse correlation length $\epsilon_1=2/\xi$ for the $5-\bar{5}$ SU(5) chain, , according to Eq.~\eqref{eq:marek}. The extrapolated value is very close to the exact result (yellow star).}
    \label{fig:xi_55bar}
\end{figure}

We have performed simulations for these systems for $N=3$ and $5$ using a unit-cell of two sites for the MPS. The correlation length can be obtained using Eq.~\eqref{eq:marek} from $2/\xi=\epsilon_1$.  Numerical data are shown in Figs.~\ref{fig:xi_33bar} and Figs.~\ref{fig:xi_55bar}. Quite remarkably, when performing extrapolations with respect to the next gap (see Fig.~\ref{fig:extrapolation} and related text), our results are in excellent agreement with the exact ones.

\begin{thebibliography}{77}%
\makeatletter
\providecommand \@ifxundefined [1]{%
 \@ifx{#1\undefined}
}%
\providecommand \@ifnum [1]{%
 \ifnum #1\expandafter \@firstoftwo
 \else \expandafter \@secondoftwo
 \fi
}%
\providecommand \@ifx [1]{%
 \ifx #1\expandafter \@firstoftwo
 \else \expandafter \@secondoftwo
 \fi
}%
\providecommand \natexlab [1]{#1}%
\providecommand \enquote  [1]{``#1''}%
\providecommand \bibnamefont  [1]{#1}%
\providecommand \bibfnamefont [1]{#1}%
\providecommand \citenamefont [1]{#1}%
\providecommand \href@noop [0]{\@secondoftwo}%
\providecommand \href [0]{\begingroup \@sanitize@url \@href}%
\providecommand \@href[1]{\@@startlink{#1}\@@href}%
\providecommand \@@href[1]{\endgroup#1\@@endlink}%
\providecommand \@sanitize@url [0]{\catcode `\\12\catcode `\$12\catcode
  `\&12\catcode `\#12\catcode `\^12\catcode `\_12\catcode `\%12\relax}%
\providecommand \@@startlink[1]{}%
\providecommand \@@endlink[0]{}%
\providecommand \url  [0]{\begingroup\@sanitize@url \@url }%
\providecommand \@url [1]{\endgroup\@href {#1}{\urlprefix }}%
\providecommand \urlprefix  [0]{URL }%
\providecommand \Eprint [0]{\href }%
\providecommand \doibase [0]{https://doi.org/}%
\providecommand \selectlanguage [0]{\@gobble}%
\providecommand \bibinfo  [0]{\@secondoftwo}%
\providecommand \bibfield  [0]{\@secondoftwo}%
\providecommand \translation [1]{[#1]}%
\providecommand \BibitemOpen [0]{}%
\providecommand \bibitemStop [0]{}%
\providecommand \bibitemNoStop [0]{.\EOS\space}%
\providecommand \EOS [0]{\spacefactor3000\relax}%
\providecommand \BibitemShut  [1]{\csname bibitem#1\endcsname}%
\let\auto@bib@innerbib\@empty
\bibitem [{\citenamefont {Herbut}(2007)}]{Herbut-book-07}%
  \BibitemOpen
  \bibfield  {author} {\bibinfo {author} {\bibfnamefont {I.}~\bibnamefont
  {Herbut}},\ }\href {https://doi.org/10.1017/CBO9780511755521} {\emph
  {\bibinfo {title} {A Modern Approach to Critical Phenomena}}}\ (\bibinfo
  {publisher} {Cambridge University Press},\ \bibinfo {year}
  {2007})\BibitemShut {NoStop}%
\bibitem [{\citenamefont {Wen}\ and\ \citenamefont {Wu}(1993)}]{Wen-W-93}%
  \BibitemOpen
  \bibfield  {author} {\bibinfo {author} {\bibfnamefont {X.-G.}\ \bibnamefont
  {Wen}}\ and\ \bibinfo {author} {\bibfnamefont {Y.-S.}\ \bibnamefont {Wu}},\
  }\bibfield  {title} {\bibinfo {title} {Transitions between the quantum hall
  states and insulators induced by periodic potentials},\ }\href
  {https://doi.org/10.1103/PhysRevLett.70.1501} {\bibfield  {journal} {\bibinfo
   {journal} {Phys. Rev. Lett.}\ }\textbf {\bibinfo {volume} {70}},\ \bibinfo
  {pages} {1501} (\bibinfo {year} {1993})}\BibitemShut {NoStop}%
\bibitem [{\citenamefont {Wen}(2000)}]{Wen-20}%
  \BibitemOpen
  \bibfield  {author} {\bibinfo {author} {\bibfnamefont {X.-G.}\ \bibnamefont
  {Wen}},\ }\bibfield  {title} {\bibinfo {title} {Continuous topological phase
  transitions between clean quantum hall states},\ }\href
  {https://doi.org/10.1103/PhysRevLett.84.3950} {\bibfield  {journal} {\bibinfo
   {journal} {Phys. Rev. Lett.}\ }\textbf {\bibinfo {volume} {84}},\ \bibinfo
  {pages} {3950} (\bibinfo {year} {2000})}\BibitemShut {NoStop}%
\bibitem [{\citenamefont {Wen}(2017)}]{Wen-RMP-17}%
  \BibitemOpen
  \bibfield  {author} {\bibinfo {author} {\bibfnamefont {X.-G.}\ \bibnamefont
  {Wen}},\ }\bibfield  {title} {\bibinfo {title} {Colloquium: Zoo of
  quantum-topological phases of matter},\ }\href
  {https://doi.org/10.1103/RevModPhys.89.041004} {\bibfield  {journal}
  {\bibinfo  {journal} {Rev. Mod. Phys.}\ }\textbf {\bibinfo {volume} {89}},\
  \bibinfo {pages} {041004} (\bibinfo {year} {2017})}\BibitemShut {NoStop}%
\bibitem [{\citenamefont {Senthil}\ \emph
  {et~al.}(2004{\natexlab{a}})\citenamefont {Senthil}, \citenamefont
  {Vishwanath}, \citenamefont {Balents}, \citenamefont {Sachdev},\ and\
  \citenamefont {Fisher}}]{Senthil-V-B-S-F-04}%
  \BibitemOpen
  \bibfield  {author} {\bibinfo {author} {\bibfnamefont {T.}~\bibnamefont
  {Senthil}}, \bibinfo {author} {\bibfnamefont {A.}~\bibnamefont {Vishwanath}},
  \bibinfo {author} {\bibfnamefont {L.}~\bibnamefont {Balents}}, \bibinfo
  {author} {\bibfnamefont {S.}~\bibnamefont {Sachdev}},\ and\ \bibinfo {author}
  {\bibfnamefont {M.~P.~A.}\ \bibnamefont {Fisher}},\ }\bibfield  {title}
  {\bibinfo {title} {{D}econfined {Q}uantum {C}ritical {P}oints},\ }\href
  {http://www.sciencemag.org/cgi/content/abstract/303/5663/ 1490} {\bibfield
  {journal} {\bibinfo  {journal} {Science}\ }\textbf {\bibinfo {volume}
  {303}},\ \bibinfo {pages} {1490} (\bibinfo {year}
  {2004}{\natexlab{a}})}\BibitemShut {NoStop}%
\bibitem [{\citenamefont {Senthil}\ \emph
  {et~al.}(2004{\natexlab{b}})\citenamefont {Senthil}, \citenamefont {Balents},
  \citenamefont {Sachdev}, \citenamefont {Vishwanath},\ and\ \citenamefont
  {Fisher}}]{Senthil-B-S-V-F-04}%
  \BibitemOpen
  \bibfield  {author} {\bibinfo {author} {\bibfnamefont {T.}~\bibnamefont
  {Senthil}}, \bibinfo {author} {\bibfnamefont {L.}~\bibnamefont {Balents}},
  \bibinfo {author} {\bibfnamefont {S.}~\bibnamefont {Sachdev}}, \bibinfo
  {author} {\bibfnamefont {A.}~\bibnamefont {Vishwanath}},\ and\ \bibinfo
  {author} {\bibfnamefont {M.~P.~A.}\ \bibnamefont {Fisher}},\ }\bibfield
  {title} {\bibinfo {title} {{Q}uantum criticality beyond the
  {L}andau-{G}inzburg-{W}ilson paradigm},\ }\href
  {http://link.aps.org/abstract/PRB/v70/e144407} {\bibfield  {journal}
  {\bibinfo  {journal} {Phys. Rev. B}\ }\textbf {\bibinfo {volume} {70}},\
  \bibinfo {pages} {144407} (\bibinfo {year} {2004}{\natexlab{b}})}\BibitemShut
  {NoStop}%
\bibitem [{\citenamefont {Senthil}(2023)}]{Senthil-23}%
  \BibitemOpen
  \bibfield  {author} {\bibinfo {author} {\bibfnamefont {T.}~\bibnamefont
  {Senthil}},\ }\href@noop {} {\bibinfo {title} {Deconfined quantum critical
  points: a review}} (\bibinfo {year} {2023}),\ \Eprint
  {https://arxiv.org/abs/2306.12638} {arXiv:2306.12638} \BibitemShut {NoStop}%
\bibitem [{\citenamefont {Zayed}\ \emph {et~al.}(2017)\citenamefont {Zayed},
  \citenamefont {R{\"u}egg}, \citenamefont {Larrea~J}, \citenamefont
  {L{\"a}uchli}, \citenamefont {Panagopoulos}, \citenamefont {Saxena},
  \citenamefont {Ellerby}, \citenamefont {McMorrow}, \citenamefont
  {Str{\"a}ssle}, \citenamefont {Klotz} \emph {et~al.}}]{Zayed2017}%
  \BibitemOpen
  \bibfield  {author} {\bibinfo {author} {\bibfnamefont {M.~E.}\ \bibnamefont
  {Zayed}}, \bibinfo {author} {\bibfnamefont {C.}~\bibnamefont {R{\"u}egg}},
  \bibinfo {author} {\bibfnamefont {J.}~\bibnamefont {Larrea~J}}, \bibinfo
  {author} {\bibfnamefont {A.}~\bibnamefont {L{\"a}uchli}}, \bibinfo {author}
  {\bibfnamefont {C.}~\bibnamefont {Panagopoulos}}, \bibinfo {author}
  {\bibfnamefont {S.}~\bibnamefont {Saxena}}, \bibinfo {author} {\bibfnamefont
  {M.}~\bibnamefont {Ellerby}}, \bibinfo {author} {\bibfnamefont
  {D.}~\bibnamefont {McMorrow}}, \bibinfo {author} {\bibfnamefont
  {T.}~\bibnamefont {Str{\"a}ssle}}, \bibinfo {author} {\bibfnamefont
  {S.}~\bibnamefont {Klotz}}, \emph {et~al.},\ }\bibfield  {title} {\bibinfo
  {title} {4-spin plaquette singlet state in the {Shastry-Sutherland} compound
  $\mathrm{SrCu}_2(\mathrm{BO}_3)_2$},\ }\href
  {https://doi.org/https://doi.org/10.1038/nphys4190} {\bibfield  {journal}
  {\bibinfo  {journal} {Nature physics}\ }\textbf {\bibinfo {volume} {13}},\
  \bibinfo {pages} {962} (\bibinfo {year} {2017})}\BibitemShut {NoStop}%
\bibitem [{\citenamefont {Guo}\ \emph {et~al.}(2020)\citenamefont {Guo},
  \citenamefont {Sun}, \citenamefont {Zhao}, \citenamefont {Wang},
  \citenamefont {Hong}, \citenamefont {Sidorov}, \citenamefont {Ma},
  \citenamefont {Wu}, \citenamefont {Li}, \citenamefont {Meng}, \citenamefont
  {Sandvik},\ and\ \citenamefont {Sun}}]{Guo2020}%
  \BibitemOpen
  \bibfield  {author} {\bibinfo {author} {\bibfnamefont {J.}~\bibnamefont
  {Guo}}, \bibinfo {author} {\bibfnamefont {G.}~\bibnamefont {Sun}}, \bibinfo
  {author} {\bibfnamefont {B.}~\bibnamefont {Zhao}}, \bibinfo {author}
  {\bibfnamefont {L.}~\bibnamefont {Wang}}, \bibinfo {author} {\bibfnamefont
  {W.}~\bibnamefont {Hong}}, \bibinfo {author} {\bibfnamefont {V.~A.}\
  \bibnamefont {Sidorov}}, \bibinfo {author} {\bibfnamefont {N.}~\bibnamefont
  {Ma}}, \bibinfo {author} {\bibfnamefont {Q.}~\bibnamefont {Wu}}, \bibinfo
  {author} {\bibfnamefont {S.}~\bibnamefont {Li}}, \bibinfo {author}
  {\bibfnamefont {Z.~Y.}\ \bibnamefont {Meng}}, \bibinfo {author}
  {\bibfnamefont {A.~W.}\ \bibnamefont {Sandvik}},\ and\ \bibinfo {author}
  {\bibfnamefont {L.}~\bibnamefont {Sun}},\ }\bibfield  {title} {\bibinfo
  {title} {Quantum phases of $\mathrm{SrCu}_2(\mathrm{BO}_3)_2$ from
  high-pressure thermodynamics},\ }\href
  {https://doi.org/10.1103/PhysRevLett.124.206602} {\bibfield  {journal}
  {\bibinfo  {journal} {Phys. Rev. Lett.}\ }\textbf {\bibinfo {volume} {124}},\
  \bibinfo {pages} {206602} (\bibinfo {year} {2020})}\BibitemShut {NoStop}%
\bibitem [{\citenamefont {Cui}\ \emph {et~al.}(2023)\citenamefont {Cui},
  \citenamefont {Liu}, \citenamefont {Lin}, \citenamefont {Wu}, \citenamefont
  {Hong}, \citenamefont {Liu}, \citenamefont {Li}, \citenamefont {Hu},
  \citenamefont {Xi}, \citenamefont {Li} \emph {et~al.}}]{Cui2023}%
  \BibitemOpen
  \bibfield  {author} {\bibinfo {author} {\bibfnamefont {Y.}~\bibnamefont
  {Cui}}, \bibinfo {author} {\bibfnamefont {L.}~\bibnamefont {Liu}}, \bibinfo
  {author} {\bibfnamefont {H.}~\bibnamefont {Lin}}, \bibinfo {author}
  {\bibfnamefont {K.-H.}\ \bibnamefont {Wu}}, \bibinfo {author} {\bibfnamefont
  {W.}~\bibnamefont {Hong}}, \bibinfo {author} {\bibfnamefont {X.}~\bibnamefont
  {Liu}}, \bibinfo {author} {\bibfnamefont {C.}~\bibnamefont {Li}}, \bibinfo
  {author} {\bibfnamefont {Z.}~\bibnamefont {Hu}}, \bibinfo {author}
  {\bibfnamefont {N.}~\bibnamefont {Xi}}, \bibinfo {author} {\bibfnamefont
  {S.}~\bibnamefont {Li}}, \emph {et~al.},\ }\bibfield  {title} {\bibinfo
  {title} {Proximate deconfined quantum critical point in
  $\mathrm{SrCu}_2(\mathrm{BO}_3)_2$},\ }\href
  {https://doi.org/10.1126/science.adc9487} {\bibfield  {journal} {\bibinfo
  {journal} {Science}\ }\textbf {\bibinfo {volume} {380}},\ \bibinfo {pages}
  {1179} (\bibinfo {year} {2023})}\BibitemShut {NoStop}%
\bibitem [{\citenamefont {Guo}\ \emph {et~al.}(2023)\citenamefont {Guo},
  \citenamefont {Wang}, \citenamefont {Huang}, \citenamefont {Chen},
  \citenamefont {Hong}, \citenamefont {Cai}, \citenamefont {Zhao},
  \citenamefont {Han}, \citenamefont {Chen}, \citenamefont {Zhou},
  \citenamefont {Li}, \citenamefont {Wu}, \citenamefont {Meng},\ and\
  \citenamefont {Sun}}]{Guo2023}%
  \BibitemOpen
  \bibfield  {author} {\bibinfo {author} {\bibfnamefont {J.}~\bibnamefont
  {Guo}}, \bibinfo {author} {\bibfnamefont {P.}~\bibnamefont {Wang}}, \bibinfo
  {author} {\bibfnamefont {C.}~\bibnamefont {Huang}}, \bibinfo {author}
  {\bibfnamefont {B.-B.}\ \bibnamefont {Chen}}, \bibinfo {author}
  {\bibfnamefont {W.}~\bibnamefont {Hong}}, \bibinfo {author} {\bibfnamefont
  {S.}~\bibnamefont {Cai}}, \bibinfo {author} {\bibfnamefont {J.}~\bibnamefont
  {Zhao}}, \bibinfo {author} {\bibfnamefont {J.}~\bibnamefont {Han}}, \bibinfo
  {author} {\bibfnamefont {X.}~\bibnamefont {Chen}}, \bibinfo {author}
  {\bibfnamefont {Y.}~\bibnamefont {Zhou}}, \bibinfo {author} {\bibfnamefont
  {S.}~\bibnamefont {Li}}, \bibinfo {author} {\bibfnamefont {Q.}~\bibnamefont
  {Wu}}, \bibinfo {author} {\bibfnamefont {Z.~Y.}\ \bibnamefont {Meng}},\ and\
  \bibinfo {author} {\bibfnamefont {L.}~\bibnamefont {Sun}},\ }\href@noop {}
  {\bibinfo {title} {Deconfined quantum critical point lost in pressurized
  $\mathrm{SrCu}_2(\mathrm{BO}_3)_2$}} (\bibinfo {year} {2023}),\ \Eprint
  {https://arxiv.org/abs/2310.20128} {arXiv:2310.20128} \BibitemShut {NoStop}%
\bibitem [{\citenamefont {Tsui}\ \emph {et~al.}(2017)\citenamefont {Tsui},
  \citenamefont {Huang}, \citenamefont {Jiang},\ and\ \citenamefont
  {Lee}}]{Tsui2017}%
  \BibitemOpen
  \bibfield  {author} {\bibinfo {author} {\bibfnamefont {L.}~\bibnamefont
  {Tsui}}, \bibinfo {author} {\bibfnamefont {Y.-T.}\ \bibnamefont {Huang}},
  \bibinfo {author} {\bibfnamefont {H.-C.}\ \bibnamefont {Jiang}},\ and\
  \bibinfo {author} {\bibfnamefont {D.-H.}\ \bibnamefont {Lee}},\ }\bibfield
  {title} {\bibinfo {title} {The phase transitions between $\mathrm{Z}_n \times
  \mathrm{Z}_n$ bosonic topological phases in 1+1{D}, and a constraint on the
  central charge for the critical points between bosonic symmetry protected
  topological phases},\ }\href
  {https://doi.org/https://doi.org/10.1016/j.nuclphysb.2017.03.021} {\bibfield
  {journal} {\bibinfo  {journal} {Nuclear Physics B}\ }\textbf {\bibinfo
  {volume} {919}},\ \bibinfo {pages} {470} (\bibinfo {year}
  {2017})}\BibitemShut {NoStop}%
\bibitem [{\citenamefont {Mudry}\ \emph {et~al.}(2019)\citenamefont {Mudry},
  \citenamefont {Furusaki}, \citenamefont {Morimoto},\ and\ \citenamefont
  {Hikihara}}]{Mudry-F-M-T-H-19}%
  \BibitemOpen
  \bibfield  {author} {\bibinfo {author} {\bibfnamefont {C.}~\bibnamefont
  {Mudry}}, \bibinfo {author} {\bibfnamefont {A.}~\bibnamefont {Furusaki}},
  \bibinfo {author} {\bibfnamefont {T.}~\bibnamefont {Morimoto}},\ and\
  \bibinfo {author} {\bibfnamefont {T.}~\bibnamefont {Hikihara}},\ }\bibfield
  {title} {\bibinfo {title} {{Quantum phase transitions beyond Landau-Ginzburg
  theory in one-dimensional space revisited}},\ }\href
  {https://doi.org/10.1103/PhysRevB.99.205153} {\bibfield  {journal} {\bibinfo
  {journal} {Phys. Rev. B}\ }\textbf {\bibinfo {volume} {99}},\ \bibinfo
  {pages} {205153} (\bibinfo {year} {2019})}\BibitemShut {NoStop}%
\bibitem [{\citenamefont {Jiang}\ and\ \citenamefont
  {Motrunich}(2019)}]{Jiang-M-19}%
  \BibitemOpen
  \bibfield  {author} {\bibinfo {author} {\bibfnamefont {S.}~\bibnamefont
  {Jiang}}\ and\ \bibinfo {author} {\bibfnamefont {O.}~\bibnamefont
  {Motrunich}},\ }\bibfield  {title} {\bibinfo {title} {Ising ferromagnet to
  valence bond solid transition in a one-dimensional spin chain: Analogies to
  deconfined quantum critical points},\ }\href
  {https://doi.org/10.1103/PhysRevB.99.075103} {\bibfield  {journal} {\bibinfo
  {journal} {Phys. Rev. B}\ }\textbf {\bibinfo {volume} {99}},\ \bibinfo
  {pages} {075103} (\bibinfo {year} {2019})}\BibitemShut {NoStop}%
\bibitem [{\citenamefont {Roberts}\ \emph {et~al.}(2019)\citenamefont
  {Roberts}, \citenamefont {Jiang},\ and\ \citenamefont
  {Motrunich}}]{Roberts-J-M-19}%
  \BibitemOpen
  \bibfield  {author} {\bibinfo {author} {\bibfnamefont {B.}~\bibnamefont
  {Roberts}}, \bibinfo {author} {\bibfnamefont {S.}~\bibnamefont {Jiang}},\
  and\ \bibinfo {author} {\bibfnamefont {O.~I.}\ \bibnamefont {Motrunich}},\
  }\bibfield  {title} {\bibinfo {title} {Deconfined quantum critical point in
  one dimension},\ }\href {https://doi.org/10.1103/PhysRevB.99.165143}
  {\bibfield  {journal} {\bibinfo  {journal} {Phys. Rev. B}\ }\textbf {\bibinfo
  {volume} {99}},\ \bibinfo {pages} {165143} (\bibinfo {year}
  {2019})}\BibitemShut {NoStop}%
\bibitem [{\citenamefont {Huang}\ \emph {et~al.}(2019)\citenamefont {Huang},
  \citenamefont {Lu}, \citenamefont {You}, \citenamefont {Meng},\ and\
  \citenamefont {Xiang}}]{Huang-LYMX-19}%
  \BibitemOpen
  \bibfield  {author} {\bibinfo {author} {\bibfnamefont {R.-Z.}\ \bibnamefont
  {Huang}}, \bibinfo {author} {\bibfnamefont {D.-C.}\ \bibnamefont {Lu}},
  \bibinfo {author} {\bibfnamefont {Y.-Z.}\ \bibnamefont {You}}, \bibinfo
  {author} {\bibfnamefont {Z.~Y.}\ \bibnamefont {Meng}},\ and\ \bibinfo
  {author} {\bibfnamefont {T.}~\bibnamefont {Xiang}},\ }\bibfield  {title}
  {\bibinfo {title} {Emergent symmetry and conserved current at a
  one-dimensional incarnation of deconfined quantum critical point},\ }\href
  {https://doi.org/10.1103/PhysRevB.100.125137} {\bibfield  {journal} {\bibinfo
   {journal} {Phys. Rev. B}\ }\textbf {\bibinfo {volume} {100}},\ \bibinfo
  {pages} {125137} (\bibinfo {year} {2019})}\BibitemShut {NoStop}%
\bibitem [{\citenamefont {Huang}\ and\ \citenamefont {Yin}(2020)}]{Huang-Y-20}%
  \BibitemOpen
  \bibfield  {author} {\bibinfo {author} {\bibfnamefont {R.-Z.}\ \bibnamefont
  {Huang}}\ and\ \bibinfo {author} {\bibfnamefont {S.}~\bibnamefont {Yin}},\
  }\bibfield  {title} {\bibinfo {title} {{Kibble-Zurek mechanism for a
  one-dimensional incarnation of a deconfined quantum critical point}},\ }\href
  {https://doi.org/10.1103/PhysRevResearch.2.023175} {\bibfield  {journal}
  {\bibinfo  {journal} {Phys. Rev. Res.}\ }\textbf {\bibinfo {volume} {2}},\
  \bibinfo {pages} {023175} (\bibinfo {year} {2020})}\BibitemShut {NoStop}%
\bibitem [{\citenamefont {Yang}\ \emph {et~al.}(2020)\citenamefont {Yang},
  \citenamefont {Yao},\ and\ \citenamefont {Sandvik}}]{Yang-Y-S-20}%
  \BibitemOpen
  \bibfield  {author} {\bibinfo {author} {\bibfnamefont {S.}~\bibnamefont
  {Yang}}, \bibinfo {author} {\bibfnamefont {D.-X.}\ \bibnamefont {Yao}},\ and\
  \bibinfo {author} {\bibfnamefont {A.~W.}\ \bibnamefont {Sandvik}},\
  }\href@noop {} {\bibinfo {title} {Deconfined quantum criticality in spin-1/2
  chains with long-range interactions}} (\bibinfo {year} {2020}),\ \Eprint
  {https://arxiv.org/abs/2001.02821} {arXiv:2001.02821} \BibitemShut {NoStop}%
\bibitem [{\citenamefont {Roberts}\ \emph {et~al.}(2021)\citenamefont
  {Roberts}, \citenamefont {Jiang},\ and\ \citenamefont
  {Motrunich}}]{Brenden-M-21}%
  \BibitemOpen
  \bibfield  {author} {\bibinfo {author} {\bibfnamefont {B.}~\bibnamefont
  {Roberts}}, \bibinfo {author} {\bibfnamefont {S.}~\bibnamefont {Jiang}},\
  and\ \bibinfo {author} {\bibfnamefont {O.~I.}\ \bibnamefont {Motrunich}},\
  }\bibfield  {title} {\bibinfo {title} {One-dimensional model for deconfined
  criticality with
  ${\mathbb{z}}_{3}\ifmmode\times\else\texttimes\fi{}{\mathbb{z}}_{3}$
  symmetry},\ }\href {https://doi.org/10.1103/PhysRevB.103.155143} {\bibfield
  {journal} {\bibinfo  {journal} {Phys. Rev. B}\ }\textbf {\bibinfo {volume}
  {103}},\ \bibinfo {pages} {155143} (\bibinfo {year} {2021})}\BibitemShut
  {NoStop}%
\bibitem [{\citenamefont {Zheng}\ \emph {et~al.}(2022)\citenamefont {Zheng},
  \citenamefont {Sheng},\ and\ \citenamefont {Lu}}]{Zheng-S-L-22}%
  \BibitemOpen
  \bibfield  {author} {\bibinfo {author} {\bibfnamefont {W.}~\bibnamefont
  {Zheng}}, \bibinfo {author} {\bibfnamefont {D.~N.}\ \bibnamefont {Sheng}},\
  and\ \bibinfo {author} {\bibfnamefont {Y.-M.}\ \bibnamefont {Lu}},\
  }\bibfield  {title} {\bibinfo {title} {{Unconventional quantum phase
  transitions in a one-dimensional Lieb-Schultz-Mattis system}},\ }\href
  {https://doi.org/10.1103/PhysRevB.105.075147} {\bibfield  {journal} {\bibinfo
   {journal} {Phys. Rev. B}\ }\textbf {\bibinfo {volume} {105}},\ \bibinfo
  {pages} {075147} (\bibinfo {year} {2022})}\BibitemShut {NoStop}%
\bibitem [{\citenamefont {Zhang}\ and\ \citenamefont
  {Levin}(2023)}]{Zhang-Levin-23}%
  \BibitemOpen
  \bibfield  {author} {\bibinfo {author} {\bibfnamefont {C.}~\bibnamefont
  {Zhang}}\ and\ \bibinfo {author} {\bibfnamefont {M.}~\bibnamefont {Levin}},\
  }\bibfield  {title} {\bibinfo {title} {Exactly solvable model for a
  deconfined quantum critical point in 1d},\ }\href
  {https://doi.org/10.1103/PhysRevLett.130.026801} {\bibfield  {journal}
  {\bibinfo  {journal} {Phys. Rev. Lett.}\ }\textbf {\bibinfo {volume} {130}},\
  \bibinfo {pages} {026801} (\bibinfo {year} {2023})}\BibitemShut {NoStop}%
\bibitem [{\citenamefont {Yang}\ \emph {et~al.}(2023)\citenamefont {Yang},
  \citenamefont {Pan}, \citenamefont {Lu},\ and\ \citenamefont
  {Yu}}]{Yang-P-L-Y-23}%
  \BibitemOpen
  \bibfield  {author} {\bibinfo {author} {\bibfnamefont {S.}~\bibnamefont
  {Yang}}, \bibinfo {author} {\bibfnamefont {Z.}~\bibnamefont {Pan}}, \bibinfo
  {author} {\bibfnamefont {D.-C.}\ \bibnamefont {Lu}},\ and\ \bibinfo {author}
  {\bibfnamefont {X.-J.}\ \bibnamefont {Yu}},\ }\bibfield  {title} {\bibinfo
  {title} {Emergent self-duality in a long-range critical spin chain: From
  deconfined criticality to first-order transition},\ }\href
  {https://doi.org/10.1103/PhysRevB.108.245152} {\bibfield  {journal} {\bibinfo
   {journal} {Phys. Rev. B}\ }\textbf {\bibinfo {volume} {108}},\ \bibinfo
  {pages} {245152} (\bibinfo {year} {2023})}\BibitemShut {NoStop}%
\bibitem [{\citenamefont {Lee}\ \emph {et~al.}(2023)\citenamefont {Lee},
  \citenamefont {Ramette}, \citenamefont {Metlitski}, \citenamefont
  {Vuleti\ifmmode~\acute{c}\else \'{c}\fi{}}, \citenamefont {Ho},\ and\
  \citenamefont {Choi}}]{Lee-R-M-V-H-C-23}%
  \BibitemOpen
  \bibfield  {author} {\bibinfo {author} {\bibfnamefont {J.~Y.}\ \bibnamefont
  {Lee}}, \bibinfo {author} {\bibfnamefont {J.}~\bibnamefont {Ramette}},
  \bibinfo {author} {\bibfnamefont {M.~A.}\ \bibnamefont {Metlitski}}, \bibinfo
  {author} {\bibfnamefont {V.}~\bibnamefont {Vuleti\ifmmode~\acute{c}\else
  \'{c}\fi{}}}, \bibinfo {author} {\bibfnamefont {W.~W.}\ \bibnamefont {Ho}},\
  and\ \bibinfo {author} {\bibfnamefont {S.}~\bibnamefont {Choi}},\ }\bibfield
  {title} {\bibinfo {title} {Landau-forbidden quantum criticality in {R}ydberg
  quantum simulators},\ }\href {https://doi.org/10.1103/PhysRevLett.131.083601}
  {\bibfield  {journal} {\bibinfo  {journal} {Phys. Rev. Lett.}\ }\textbf
  {\bibinfo {volume} {131}},\ \bibinfo {pages} {083601} (\bibinfo {year}
  {2023})}\BibitemShut {NoStop}%
\bibitem [{\citenamefont {Romen}\ \emph {et~al.}(2024)\citenamefont {Romen},
  \citenamefont {Birnkammer},\ and\ \citenamefont {Knap}}]{Romen-B-K-24}%
  \BibitemOpen
  \bibfield  {author} {\bibinfo {author} {\bibfnamefont {A.}~\bibnamefont
  {Romen}}, \bibinfo {author} {\bibfnamefont {S.}~\bibnamefont {Birnkammer}},\
  and\ \bibinfo {author} {\bibfnamefont {M.}~\bibnamefont {Knap}},\ }\bibfield
  {title} {\bibinfo {title} {{Deconfined quantum criticality in the long-range,
  anisotropic Heisenberg chain}},\ }\href
  {https://doi.org/10.21468/SciPostPhysCore.7.1.008} {\bibfield  {journal}
  {\bibinfo  {journal} {SciPost Phys. Core}\ }\textbf {\bibinfo {volume} {7}},\
  \bibinfo {pages} {008} (\bibinfo {year} {2024})}\BibitemShut {NoStop}%
\bibitem [{\citenamefont {Haldane}(1981)}]{Haldane-JPC-81}%
  \BibitemOpen
  \bibfield  {author} {\bibinfo {author} {\bibfnamefont {F.~D.~M.}\
  \bibnamefont {Haldane}},\ }\bibfield  {title} {\bibinfo {title} {{'Luttinger
  liquid theory' of one-dimensional quantum fluids. I. Properties of the
  Luttinger model and their extension to the general 1D interacting spinless
  Fermi gas}},\ }\href {http://stacks.iop.org/0022-3719/14/i=19/a=010}
  {\bibfield  {journal} {\bibinfo  {journal} {Journal of Physics C: Solid State
  Physics}\ }\textbf {\bibinfo {volume} {14}},\ \bibinfo {pages} {2585}
  (\bibinfo {year} {1981})}\BibitemShut {NoStop}%
\bibitem [{\citenamefont {Hida}(1992)}]{Hida-92a}%
  \BibitemOpen
  \bibfield  {author} {\bibinfo {author} {\bibfnamefont {K.}~\bibnamefont
  {Hida}},\ }\bibfield  {title} {\bibinfo {title} {{Crossover between the
  Haldane-gap phase and the dimer phase in the spin-1/2 alternating Heisenberg
  chain}},\ }\href {http://link.aps.org/doi/10.1103/PhysRevB.45.2207}
  {\bibfield  {journal} {\bibinfo  {journal} {Phys. Rev. B}\ }\textbf {\bibinfo
  {volume} {45}},\ \bibinfo {pages} {2207} (\bibinfo {year}
  {1992})}\BibitemShut {NoStop}%
\bibitem [{\citenamefont {Haldane}(2017)}]{Haldane-Novel-lec-17}%
  \BibitemOpen
  \bibfield  {author} {\bibinfo {author} {\bibfnamefont {F.~D.~M.}\
  \bibnamefont {Haldane}},\ }\bibfield  {title} {\bibinfo {title} {Nobel
  lecture: Topological quantum matter},\ }\href
  {https://doi.org/10.1103/RevModPhys.89.040502} {\bibfield  {journal}
  {\bibinfo  {journal} {Rev. Mod. Phys.}\ }\textbf {\bibinfo {volume} {89}},\
  \bibinfo {pages} {040502} (\bibinfo {year} {2017})}\BibitemShut {NoStop}%
\bibitem [{\citenamefont {Bi}\ \emph {et~al.}(2020)\citenamefont {Bi},
  \citenamefont {Lake},\ and\ \citenamefont {Senthil}}]{Bi-L-S-20}%
  \BibitemOpen
  \bibfield  {author} {\bibinfo {author} {\bibfnamefont {Z.}~\bibnamefont
  {Bi}}, \bibinfo {author} {\bibfnamefont {E.}~\bibnamefont {Lake}},\ and\
  \bibinfo {author} {\bibfnamefont {T.}~\bibnamefont {Senthil}},\ }\bibfield
  {title} {\bibinfo {title} {{Landau ordering phase transitions beyond the
  Landau paradigm}},\ }\href {https://doi.org/10.1103/PhysRevResearch.2.023031}
  {\bibfield  {journal} {\bibinfo  {journal} {Phys. Rev. Res.}\ }\textbf
  {\bibinfo {volume} {2}},\ \bibinfo {pages} {023031} (\bibinfo {year}
  {2020})}\BibitemShut {NoStop}%
\bibitem [{\citenamefont {Duivenvoorden}\ and\ \citenamefont
  {Quella}(2013)}]{Duivenvoorden-Q-13}%
  \BibitemOpen
  \bibfield  {author} {\bibinfo {author} {\bibfnamefont {K.}~\bibnamefont
  {Duivenvoorden}}\ and\ \bibinfo {author} {\bibfnamefont {T.}~\bibnamefont
  {Quella}},\ }\bibfield  {title} {\bibinfo {title} {Topological phases of spin
  chains},\ }\href {http://link.aps.org/doi/10.1103/PhysRevB.87.125145}
  {\bibfield  {journal} {\bibinfo  {journal} {Phys. Rev. B}\ }\textbf {\bibinfo
  {volume} {87}},\ \bibinfo {pages} {125145} (\bibinfo {year}
  {2013})}\BibitemShut {NoStop}%
\bibitem [{\citenamefont {Affleck}(1986)}]{Affleck-NP86}%
  \BibitemOpen
  \bibfield  {author} {\bibinfo {author} {\bibfnamefont {I.}~\bibnamefont
  {Affleck}},\ }\bibfield  {title} {\bibinfo {title} {Exact critical exponents
  for quantum spin chains, non-linear $\sigma$-models at $\theta=\pi$ and the
  quantum {H}all effect},\ }\href
  {https://doi.org/http://dx.doi.org/10.1016/0550-3213(86)90167-7} {\bibfield
  {journal} {\bibinfo  {journal} {Nuclear Physics B}\ }\textbf {\bibinfo
  {volume} {265}},\ \bibinfo {pages} {409 } (\bibinfo {year}
  {1986})}\BibitemShut {NoStop}%
\bibitem [{\citenamefont {Affleck}(1988)}]{Affleck-88}%
  \BibitemOpen
  \bibfield  {author} {\bibinfo {author} {\bibfnamefont {I.}~\bibnamefont
  {Affleck}},\ }\bibfield  {title} {\bibinfo {title} {Critical behaviour of
  su($n$) quantum chains and topological non-linear $\sigma$-models},\ }\href
  {http://www.sciencedirect.com/science/article/B6TVC-4718MMW-
  12C/2/04da87f9aa53d2aaa48342141fc9be83} {\bibfield  {journal} {\bibinfo
  {journal} {Nuclear Physics B}\ }\textbf {\bibinfo {volume} {305}},\ \bibinfo
  {pages} {582} (\bibinfo {year} {1988})}\BibitemShut {NoStop}%
\bibitem [{\citenamefont {James}\ \emph {et~al.}(2018)\citenamefont {James},
  \citenamefont {Konik}, \citenamefont {Lecheminant}, \citenamefont
  {Robinson},\ and\ \citenamefont {Tsvelik}}]{James-K-L-R-T-18}%
  \BibitemOpen
  \bibfield  {author} {\bibinfo {author} {\bibfnamefont {A.~J.~A.}\
  \bibnamefont {James}}, \bibinfo {author} {\bibfnamefont {R.~M.}\ \bibnamefont
  {Konik}}, \bibinfo {author} {\bibfnamefont {P.}~\bibnamefont {Lecheminant}},
  \bibinfo {author} {\bibfnamefont {N.~J.}\ \bibnamefont {Robinson}},\ and\
  \bibinfo {author} {\bibfnamefont {A.~M.}\ \bibnamefont {Tsvelik}},\
  }\bibfield  {title} {\bibinfo {title} {Non-perturbative methodologies for
  low-dimensional strongly-correlated systems: From non-abelian bosonization to
  truncated spectrum methods},\ }\href
  {http://stacks.iop.org/0034-4885/81/i=4/a=046002} {\bibfield  {journal}
  {\bibinfo  {journal} {Reports on Progress in Physics}\ }\textbf {\bibinfo
  {volume} {81}},\ \bibinfo {pages} {046002} (\bibinfo {year}
  {2018})}\BibitemShut {NoStop}%
\bibitem [{sup()}]{supmat}%
  \BibitemOpen
  \href@noop {} {}\bibinfo {note} {See Supplementary Material (SM) for
  technical details about the continuum limit and the numerical method, as well
  as additional numerical data.}\BibitemShut {Stop}%
\bibitem [{\citenamefont {Lajk\'{o}}\ \emph {et~al.}(2017)\citenamefont
  {Lajk\'{o}}, \citenamefont {Wamer}, \citenamefont {Mila},\ and\ \citenamefont
  {Affleck}}]{Lajko-W-M-A-17}%
  \BibitemOpen
  \bibfield  {author} {\bibinfo {author} {\bibfnamefont {M.}~\bibnamefont
  {Lajk\'{o}}}, \bibinfo {author} {\bibfnamefont {K.}~\bibnamefont {Wamer}},
  \bibinfo {author} {\bibfnamefont {F.}~\bibnamefont {Mila}},\ and\ \bibinfo
  {author} {\bibfnamefont {I.}~\bibnamefont {Affleck}},\ }\bibfield  {title}
  {\bibinfo {title} {{Generalization of the Haldane conjecture to SU(3)
  chains}},\ }\href
  {https://doi.org/https://doi.org/10.1016/j.nuclphysb.2017.09.015} {\bibfield
  {journal} {\bibinfo  {journal} {Nuclear Physics B}\ }\textbf {\bibinfo
  {volume} {924}},\ \bibinfo {pages} {508 } (\bibinfo {year}
  {2017})}\BibitemShut {NoStop}%
\bibitem [{\citenamefont {Rachel}\ \emph {et~al.}(2010)\citenamefont {Rachel},
  \citenamefont {Schuricht}, \citenamefont {Scharfenberger}, \citenamefont
  {Thomale},\ and\ \citenamefont {Greiter}}]{Rachel-S-S-T-G-10}%
  \BibitemOpen
  \bibfield  {author} {\bibinfo {author} {\bibfnamefont {S.}~\bibnamefont
  {Rachel}}, \bibinfo {author} {\bibfnamefont {D.}~\bibnamefont {Schuricht}},
  \bibinfo {author} {\bibfnamefont {B.}~\bibnamefont {Scharfenberger}},
  \bibinfo {author} {\bibfnamefont {R.}~\bibnamefont {Thomale}},\ and\ \bibinfo
  {author} {\bibfnamefont {M.}~\bibnamefont {Greiter}},\ }\bibfield  {title}
  {\bibinfo {title} {Spontaneous parity violation in a quantum spin chain},\
  }\href {http://stacks.iop.org/1742-6596/200/i=2/a=022049} {\bibfield
  {journal} {\bibinfo  {journal} {J. Phys.: Conference Series}\ }\textbf
  {\bibinfo {volume} {200}},\ \bibinfo {pages} {022049} (\bibinfo {year}
  {2010})}\BibitemShut {NoStop}%
\bibitem [{\citenamefont {Morimoto}\ \emph {et~al.}(2014)\citenamefont
  {Morimoto}, \citenamefont {Ueda}, \citenamefont {Momoi},\ and\ \citenamefont
  {Furusaki}}]{Morimoto-U-M-F-14}%
  \BibitemOpen
  \bibfield  {author} {\bibinfo {author} {\bibfnamefont {T.}~\bibnamefont
  {Morimoto}}, \bibinfo {author} {\bibfnamefont {H.}~\bibnamefont {Ueda}},
  \bibinfo {author} {\bibfnamefont {T.}~\bibnamefont {Momoi}},\ and\ \bibinfo
  {author} {\bibfnamefont {A.}~\bibnamefont {Furusaki}},\ }\bibfield  {title}
  {\bibinfo {title} {{Z$_3$} symmetry-protected topological phases in {SU}(3)
  {AKLT} model},\ }\href {https://doi.org/10.1103/PhysRevB.90.235111}
  {\bibfield  {journal} {\bibinfo  {journal} {Phys. Rev. B}\ }\textbf {\bibinfo
  {volume} {90}},\ \bibinfo {pages} {235111} (\bibinfo {year}
  {2014})}\BibitemShut {NoStop}%
\bibitem [{\citenamefont {Ueda}\ \emph {et~al.}(2018)\citenamefont {Ueda},
  \citenamefont {Morimoto},\ and\ \citenamefont {Momoi}}]{Ueda-M-M-18}%
  \BibitemOpen
  \bibfield  {author} {\bibinfo {author} {\bibfnamefont {H.}~\bibnamefont
  {Ueda}}, \bibinfo {author} {\bibfnamefont {T.}~\bibnamefont {Morimoto}},\
  and\ \bibinfo {author} {\bibfnamefont {T.}~\bibnamefont {Momoi}},\ }\bibfield
   {title} {\bibinfo {title} {Symmetry protected topological phases in
  two-orbital su(4) fermionic atoms},\ }\href
  {https://doi.org/10.1103/PhysRevB.98.045128} {\bibfield  {journal} {\bibinfo
  {journal} {Phys. Rev. B}\ }\textbf {\bibinfo {volume} {98}},\ \bibinfo
  {pages} {045128} (\bibinfo {year} {2018})}\BibitemShut {NoStop}%
\bibitem [{\citenamefont {Roy}\ and\ \citenamefont {Quella}(2018)}]{Roy-Q-18}%
  \BibitemOpen
  \bibfield  {author} {\bibinfo {author} {\bibfnamefont {A.}~\bibnamefont
  {Roy}}\ and\ \bibinfo {author} {\bibfnamefont {T.}~\bibnamefont {Quella}},\
  }\bibfield  {title} {\bibinfo {title} {Chiral haldane phases of
  $\text{SU}(n)$ quantum spin chains},\ }\href
  {https://doi.org/10.1103/PhysRevB.97.155148} {\bibfield  {journal} {\bibinfo
  {journal} {Phys. Rev. B}\ }\textbf {\bibinfo {volume} {97}},\ \bibinfo
  {pages} {155148} (\bibinfo {year} {2018})}\BibitemShut {NoStop}%
\bibitem [{\citenamefont {Fromholz}\ \emph {et~al.}(2019)\citenamefont
  {Fromholz}, \citenamefont {Capponi}, \citenamefont {Lecheminant},
  \citenamefont {Papoular},\ and\ \citenamefont
  {Totsuka}}]{Fromholz-C-L-P-T-19}%
  \BibitemOpen
  \bibfield  {author} {\bibinfo {author} {\bibfnamefont {P.}~\bibnamefont
  {Fromholz}}, \bibinfo {author} {\bibfnamefont {S.}~\bibnamefont {Capponi}},
  \bibinfo {author} {\bibfnamefont {P.}~\bibnamefont {Lecheminant}}, \bibinfo
  {author} {\bibfnamefont {D.~J.}\ \bibnamefont {Papoular}},\ and\ \bibinfo
  {author} {\bibfnamefont {K.}~\bibnamefont {Totsuka}},\ }\bibfield  {title}
  {\bibinfo {title} {Haldane phases with ultracold fermionic atoms in
  double-well optical lattices},\ }\href
  {https://doi.org/10.1103/PhysRevB.99.054414} {\bibfield  {journal} {\bibinfo
  {journal} {Phys. Rev. B}\ }\textbf {\bibinfo {volume} {99}},\ \bibinfo
  {pages} {054414} (\bibinfo {year} {2019})}\BibitemShut {NoStop}%
\bibitem [{\citenamefont {Capponi}\ \emph {et~al.}(2020)\citenamefont
  {Capponi}, \citenamefont {Fromholz}, \citenamefont {Lecheminant},\ and\
  \citenamefont {Totsuka}}]{Capponi-F-L-T-20}%
  \BibitemOpen
  \bibfield  {author} {\bibinfo {author} {\bibfnamefont {S.}~\bibnamefont
  {Capponi}}, \bibinfo {author} {\bibfnamefont {P.}~\bibnamefont {Fromholz}},
  \bibinfo {author} {\bibfnamefont {P.}~\bibnamefont {Lecheminant}},\ and\
  \bibinfo {author} {\bibfnamefont {K.}~\bibnamefont {Totsuka}},\ }\bibfield
  {title} {\bibinfo {title} {Symmetry-protected topological phases in a two-leg
  $\text{SU}(n)$ spin ladder with unequal spins},\ }\href
  {https://doi.org/10.1103/PhysRevB.101.195121} {\bibfield  {journal} {\bibinfo
   {journal} {Phys. Rev. B}\ }\textbf {\bibinfo {volume} {101}},\ \bibinfo
  {pages} {195121} (\bibinfo {year} {2020})}\BibitemShut {NoStop}%
\bibitem [{\citenamefont {Verresen}\ \emph {et~al.}(2017)\citenamefont
  {Verresen}, \citenamefont {Moessner},\ and\ \citenamefont
  {Pollmann}}]{Verresen-M-P-17}%
  \BibitemOpen
  \bibfield  {author} {\bibinfo {author} {\bibfnamefont {R.}~\bibnamefont
  {Verresen}}, \bibinfo {author} {\bibfnamefont {R.}~\bibnamefont {Moessner}},\
  and\ \bibinfo {author} {\bibfnamefont {F.}~\bibnamefont {Pollmann}},\
  }\bibfield  {title} {\bibinfo {title} {One-dimensional symmetry protected
  topological phases and their transitions},\ }\href
  {https://doi.org/10.1103/PhysRevB.96.165124} {\bibfield  {journal} {\bibinfo
  {journal} {Phys. Rev. B}\ }\textbf {\bibinfo {volume} {96}},\ \bibinfo
  {pages} {165124} (\bibinfo {year} {2017})}\BibitemShut {NoStop}%
\bibitem [{\citenamefont {Affleck}(1990)}]{Affleck-SUN-90}%
  \BibitemOpen
  \bibfield  {author} {\bibinfo {author} {\bibfnamefont {I.}~\bibnamefont
  {Affleck}},\ }\bibfield  {title} {\bibinfo {title} {Exact results on the
  dimerisation transition in {SU($n$)} antiferromagnetic chains},\ }\href
  {http://stacks.iop.org/0953-8984/2/i=2/a=016} {\bibfield  {journal} {\bibinfo
   {journal} {Journal of Physics: Condensed Matter}\ }\textbf {\bibinfo
  {volume} {2}},\ \bibinfo {pages} {405} (\bibinfo {year} {1990})}\BibitemShut
  {NoStop}%
\bibitem [{\citenamefont {Barber}\ and\ \citenamefont
  {Batchelor}(1989)}]{Barber-B-89}%
  \BibitemOpen
  \bibfield  {author} {\bibinfo {author} {\bibfnamefont {M.~N.}\ \bibnamefont
  {Barber}}\ and\ \bibinfo {author} {\bibfnamefont {M.~T.}\ \bibnamefont
  {Batchelor}},\ }\bibfield  {title} {\bibinfo {title} {Spectrum of the
  biquadratic spin-1 antiferromagnetic chain},\ }\href
  {https://doi.org/10.1103/PhysRevB.40.4621} {\bibfield  {journal} {\bibinfo
  {journal} {Phys. Rev. B}\ }\textbf {\bibinfo {volume} {40}},\ \bibinfo
  {pages} {4621} (\bibinfo {year} {1989})}\BibitemShut {NoStop}%
\bibitem [{\citenamefont {Kl\"{u}mper}(1989)}]{Klumper-89}%
  \BibitemOpen
  \bibfield  {author} {\bibinfo {author} {\bibfnamefont {A.}~\bibnamefont
  {Kl\"{u}mper}},\ }\bibfield  {title} {\bibinfo {title} {New results for
  $q$-state vertex models and the pure biquadratic spin-1 {H}amiltonian},\
  }\href {http://stacks.iop.org/0295-5075/9/i=8/a=013} {\bibfield  {journal}
  {\bibinfo  {journal} {Europhysics Letters}\ }\textbf {\bibinfo {volume}
  {9}},\ \bibinfo {pages} {815} (\bibinfo {year} {1989})}\BibitemShut {NoStop}%
\bibitem [{\citenamefont {Kl\"umper}(1990)}]{Klumper-90}%
  \BibitemOpen
  \bibfield  {author} {\bibinfo {author} {\bibfnamefont {A.}~\bibnamefont
  {Kl\"umper}},\ }\bibfield  {title} {\bibinfo {title} {The spectra of
  $q$-state vertex models and related antiferromagnetic quantum spin chains},\
  }\href {https://doi.org/10.1088/0305-4470/23/5/023} {\bibfield  {journal}
  {\bibinfo  {journal} {Journal of Physics A: Mathematical and General}\
  }\textbf {\bibinfo {volume} {23}},\ \bibinfo {pages} {809} (\bibinfo {year}
  {1990})}\BibitemShut {NoStop}%
\bibitem [{\citenamefont {Cirac}\ \emph {et~al.}(2021)\citenamefont {Cirac},
  \citenamefont {P\'erez-Garc\'{\i}a}, \citenamefont {Schuch},\ and\
  \citenamefont {Verstraete}}]{Cirac2021}%
  \BibitemOpen
  \bibfield  {author} {\bibinfo {author} {\bibfnamefont {J.~I.}\ \bibnamefont
  {Cirac}}, \bibinfo {author} {\bibfnamefont {D.}~\bibnamefont
  {P\'erez-Garc\'{\i}a}}, \bibinfo {author} {\bibfnamefont {N.}~\bibnamefont
  {Schuch}},\ and\ \bibinfo {author} {\bibfnamefont {F.}~\bibnamefont
  {Verstraete}},\ }\bibfield  {title} {\bibinfo {title} {Matrix product states
  and projected entangled pair states: Concepts, symmetries, theorems},\ }\href
  {https://doi.org/10.1103/RevModPhys.93.045003} {\bibfield  {journal}
  {\bibinfo  {journal} {Rev. Mod. Phys.}\ }\textbf {\bibinfo {volume} {93}},\
  \bibinfo {pages} {045003} (\bibinfo {year} {2021})}\BibitemShut {NoStop}%
\bibitem [{\citenamefont {Vanderstraeten}\ \emph {et~al.}(2019)\citenamefont
  {Vanderstraeten}, \citenamefont {Haegeman},\ and\ \citenamefont
  {Verstraete}}]{Vanderstraeten2019}%
  \BibitemOpen
  \bibfield  {author} {\bibinfo {author} {\bibfnamefont {L.}~\bibnamefont
  {Vanderstraeten}}, \bibinfo {author} {\bibfnamefont {J.}~\bibnamefont
  {Haegeman}},\ and\ \bibinfo {author} {\bibfnamefont {F.}~\bibnamefont
  {Verstraete}},\ }\bibfield  {title} {\bibinfo {title} {{Tangent-space methods
  for uniform matrix product states}},\ }\href
  {https://doi.org/10.21468/SciPostPhysLectNotes.7} {\bibfield  {journal}
  {\bibinfo  {journal} {SciPost Phys. Lect. Notes}\ ,\ \bibinfo {pages} {7}}
  (\bibinfo {year} {2019})}\BibitemShut {NoStop}%
\bibitem [{\citenamefont {McCulloch}(2008)}]{McCulloch2008}%
  \BibitemOpen
  \bibfield  {author} {\bibinfo {author} {\bibfnamefont {I.~P.}\ \bibnamefont
  {McCulloch}},\ }\href@noop {} {\bibinfo {title} {Infinite size density matrix
  renormalization group, revisited}} (\bibinfo {year} {2008}),\ \Eprint
  {https://arxiv.org/abs/0804.2509} {arXiv:0804.2509} \BibitemShut {NoStop}%
\bibitem [{\citenamefont {Zauner-Stauber}\ \emph {et~al.}(2018)\citenamefont
  {Zauner-Stauber}, \citenamefont {Vanderstraeten}, \citenamefont {Fishman},
  \citenamefont {Verstraete},\ and\ \citenamefont
  {Haegeman}}]{ZaunerStauber2018}%
  \BibitemOpen
  \bibfield  {author} {\bibinfo {author} {\bibfnamefont {V.}~\bibnamefont
  {Zauner-Stauber}}, \bibinfo {author} {\bibfnamefont {L.}~\bibnamefont
  {Vanderstraeten}}, \bibinfo {author} {\bibfnamefont {M.~T.}\ \bibnamefont
  {Fishman}}, \bibinfo {author} {\bibfnamefont {F.}~\bibnamefont
  {Verstraete}},\ and\ \bibinfo {author} {\bibfnamefont {J.}~\bibnamefont
  {Haegeman}},\ }\bibfield  {title} {\bibinfo {title} {Variational optimization
  algorithms for uniform matrix product states},\ }\href
  {https://doi.org/10.1103/PhysRevB.97.045145} {\bibfield  {journal} {\bibinfo
  {journal} {Phys. Rev. B}\ }\textbf {\bibinfo {volume} {97}},\ \bibinfo
  {pages} {045145} (\bibinfo {year} {2018})}\BibitemShut {NoStop}%
\bibitem [{\citenamefont {Devos}\ \emph {et~al.}(2024)\citenamefont {Devos},
  \citenamefont {Burgelman}, \citenamefont {Mortier}, \citenamefont {Vanhecke},
  \citenamefont {Haegeman}, \citenamefont {Verstraete},\ and\ \citenamefont
  {Vanderstraeten}}]{TensorTrack2024}%
  \BibitemOpen
  \bibfield  {author} {\bibinfo {author} {\bibfnamefont {L.}~\bibnamefont
  {Devos}}, \bibinfo {author} {\bibfnamefont {L.}~\bibnamefont {Burgelman}},
  \bibinfo {author} {\bibfnamefont {Q.}~\bibnamefont {Mortier}}, \bibinfo
  {author} {\bibfnamefont {B.}~\bibnamefont {Vanhecke}}, \bibinfo {author}
  {\bibfnamefont {J.}~\bibnamefont {Haegeman}}, \bibinfo {author}
  {\bibfnamefont {F.}~\bibnamefont {Verstraete}},\ and\ \bibinfo {author}
  {\bibfnamefont {L.}~\bibnamefont {Vanderstraeten}},\ }\href
  {https://doi.org/10.5281/ZENODO.6670354} {\bibinfo {title} {{{TensorTrack}}}}
  (\bibinfo {year} {2024})\BibitemShut {NoStop}%
\bibitem [{\citenamefont {Haegeman}\ \emph {et~al.}(2024)\citenamefont
  {Haegeman}, \citenamefont {Devos},\ and\ \citenamefont {{Van
  Damme}}}]{TensorKit2024}%
  \BibitemOpen
  \bibfield  {author} {\bibinfo {author} {\bibfnamefont {J.}~\bibnamefont
  {Haegeman}}, \bibinfo {author} {\bibfnamefont {L.}~\bibnamefont {Devos}},\
  and\ \bibinfo {author} {\bibfnamefont {M.}~\bibnamefont {{Van Damme}}},\
  }\href {https://doi.org/10.5281/ZENODO.8421339} {\bibinfo {title}
  {{{TensorKit}}.jl}} (\bibinfo {year} {2024})\BibitemShut {NoStop}%
\bibitem [{\citenamefont {{Van Damme}}\ and\ \citenamefont
  {Devos}(2023)}]{MPSKit2023}%
  \BibitemOpen
  \bibfield  {author} {\bibinfo {author} {\bibfnamefont {M.}~\bibnamefont {{Van
  Damme}}}\ and\ \bibinfo {author} {\bibfnamefont {L.}~\bibnamefont {Devos}},\
  }\href {https://doi.org/10.5281/ZENODO.10654900} {\bibinfo {title}
  {{{MPSKit}}.jl}} (\bibinfo {year} {2023})\BibitemShut {NoStop}%
\bibitem [{\citenamefont {Rams}\ \emph {et~al.}(2018)\citenamefont {Rams},
  \citenamefont {Czarnik},\ and\ \citenamefont {Cincio}}]{Rams2018}%
  \BibitemOpen
  \bibfield  {author} {\bibinfo {author} {\bibfnamefont {M.~M.}\ \bibnamefont
  {Rams}}, \bibinfo {author} {\bibfnamefont {P.}~\bibnamefont {Czarnik}},\ and\
  \bibinfo {author} {\bibfnamefont {L.}~\bibnamefont {Cincio}},\ }\bibfield
  {title} {\bibinfo {title} {Precise extrapolation of the correlation function
  asymptotics in uniform tensor network states with application to the
  {B}ose-{H}ubbard and {XXZ} models},\ }\href
  {https://doi.org/10.1103/PhysRevX.8.041033} {\bibfield  {journal} {\bibinfo
  {journal} {Phys. Rev. X}\ }\textbf {\bibinfo {volume} {8}},\ \bibinfo {pages}
  {041033} (\bibinfo {year} {2018})}\BibitemShut {NoStop}%
\bibitem [{\citenamefont {Vanhecke}\ \emph {et~al.}(2019)\citenamefont
  {Vanhecke}, \citenamefont {Haegeman}, \citenamefont {Van~Acoleyen},
  \citenamefont {Vanderstraeten},\ and\ \citenamefont
  {Verstraete}}]{Vanhecke2019}%
  \BibitemOpen
  \bibfield  {author} {\bibinfo {author} {\bibfnamefont {B.}~\bibnamefont
  {Vanhecke}}, \bibinfo {author} {\bibfnamefont {J.}~\bibnamefont {Haegeman}},
  \bibinfo {author} {\bibfnamefont {K.}~\bibnamefont {Van~Acoleyen}}, \bibinfo
  {author} {\bibfnamefont {L.}~\bibnamefont {Vanderstraeten}},\ and\ \bibinfo
  {author} {\bibfnamefont {F.}~\bibnamefont {Verstraete}},\ }\bibfield  {title}
  {\bibinfo {title} {Scaling hypothesis for matrix product states},\ }\href
  {https://doi.org/10.1103/PhysRevLett.123.250604} {\bibfield  {journal}
  {\bibinfo  {journal} {Phys. Rev. Lett.}\ }\textbf {\bibinfo {volume} {123}},\
  \bibinfo {pages} {250604} (\bibinfo {year} {2019})}\BibitemShut {NoStop}%
\bibitem [{\citenamefont {Su}\ and\ \citenamefont {Schrieffer}(1981)}]{Su1981}%
  \BibitemOpen
  \bibfield  {author} {\bibinfo {author} {\bibfnamefont {W.~P.}\ \bibnamefont
  {Su}}\ and\ \bibinfo {author} {\bibfnamefont {J.~R.}\ \bibnamefont
  {Schrieffer}},\ }\bibfield  {title} {\bibinfo {title} {Fractionally charged
  excitations in charge-density-wave systems with commensurability 3},\ }\href
  {http://link.aps.org/abstract/PRL/v46/p738} {\bibfield  {journal} {\bibinfo
  {journal} {Phys. Rev. Lett.}\ }\textbf {\bibinfo {volume} {46}},\ \bibinfo
  {pages} {738} (\bibinfo {year} {1981})}\BibitemShut {NoStop}%
\bibitem [{\citenamefont {Shastry}\ and\ \citenamefont
  {Sutherland}(1981)}]{Shastry1981}%
  \BibitemOpen
  \bibfield  {author} {\bibinfo {author} {\bibfnamefont {B.~S.}\ \bibnamefont
  {Shastry}}\ and\ \bibinfo {author} {\bibfnamefont {B.}~\bibnamefont
  {Sutherland}},\ }\bibfield  {title} {\bibinfo {title} {Excitation spectrum of
  a dimerized next-neighbor antiferromagnetic chain},\ }\href
  {https://doi.org/10.1103/PhysRevLett.47.964} {\bibfield  {journal} {\bibinfo
  {journal} {Phys. Rev. Lett.}\ }\textbf {\bibinfo {volume} {47}},\ \bibinfo
  {pages} {964} (\bibinfo {year} {1981})}\BibitemShut {NoStop}%
\bibitem [{\citenamefont {Faddeev}\ and\ \citenamefont
  {Takhtajan}(1981)}]{Faddeev1981}%
  \BibitemOpen
  \bibfield  {author} {\bibinfo {author} {\bibfnamefont {L.}~\bibnamefont
  {Faddeev}}\ and\ \bibinfo {author} {\bibfnamefont {L.}~\bibnamefont
  {Takhtajan}},\ }\bibfield  {title} {\bibinfo {title} {What is the spin of a
  spin wave?},\ }\href
  {https://doi.org/https://doi.org/10.1016/0375-9601(81)90335-2} {\bibfield
  {journal} {\bibinfo  {journal} {Physics Letters A}\ }\textbf {\bibinfo
  {volume} {85}},\ \bibinfo {pages} {375} (\bibinfo {year} {1981})}\BibitemShut
  {NoStop}%
\bibitem [{\citenamefont {Vanderstraeten}\ \emph {et~al.}(2020)\citenamefont
  {Vanderstraeten}, \citenamefont {Wybo}, \citenamefont {Chepiga},
  \citenamefont {Verstraete},\ and\ \citenamefont {Mila}}]{Vanderstraeten2020}%
  \BibitemOpen
  \bibfield  {author} {\bibinfo {author} {\bibfnamefont {L.}~\bibnamefont
  {Vanderstraeten}}, \bibinfo {author} {\bibfnamefont {E.}~\bibnamefont
  {Wybo}}, \bibinfo {author} {\bibfnamefont {N.}~\bibnamefont {Chepiga}},
  \bibinfo {author} {\bibfnamefont {F.}~\bibnamefont {Verstraete}},\ and\
  \bibinfo {author} {\bibfnamefont {F.}~\bibnamefont {Mila}},\ }\bibfield
  {title} {\bibinfo {title} {Spinon confinement and deconfinement in spin-1
  chains},\ }\href {https://doi.org/10.1103/PhysRevB.101.115138} {\bibfield
  {journal} {\bibinfo  {journal} {Phys. Rev. B}\ }\textbf {\bibinfo {volume}
  {101}},\ \bibinfo {pages} {115138} (\bibinfo {year} {2020})}\BibitemShut
  {NoStop}%
\bibitem [{\citenamefont {Haegeman}\ \emph {et~al.}(2012)\citenamefont
  {Haegeman}, \citenamefont {Pirvu}, \citenamefont {Weir}, \citenamefont
  {Cirac}, \citenamefont {Osborne}, \citenamefont {Verschelde},\ and\
  \citenamefont {Verstraete}}]{Haegeman2012}%
  \BibitemOpen
  \bibfield  {author} {\bibinfo {author} {\bibfnamefont {J.}~\bibnamefont
  {Haegeman}}, \bibinfo {author} {\bibfnamefont {B.}~\bibnamefont {Pirvu}},
  \bibinfo {author} {\bibfnamefont {D.~J.}\ \bibnamefont {Weir}}, \bibinfo
  {author} {\bibfnamefont {J.~I.}\ \bibnamefont {Cirac}}, \bibinfo {author}
  {\bibfnamefont {T.~J.}\ \bibnamefont {Osborne}}, \bibinfo {author}
  {\bibfnamefont {H.}~\bibnamefont {Verschelde}},\ and\ \bibinfo {author}
  {\bibfnamefont {F.}~\bibnamefont {Verstraete}},\ }\bibfield  {title}
  {\bibinfo {title} {Variational matrix product ansatz for dispersion
  relations},\ }\href {https://doi.org/10.1103/PhysRevB.85.100408} {\bibfield
  {journal} {\bibinfo  {journal} {Phys. Rev. B}\ }\textbf {\bibinfo {volume}
  {85}},\ \bibinfo {pages} {100408} (\bibinfo {year} {2012})}\BibitemShut
  {NoStop}%
\bibitem [{\citenamefont {Zaletel}\ \emph {et~al.}(2015)\citenamefont
  {Zaletel}, \citenamefont {Mong}, \citenamefont {Karrasch}, \citenamefont
  {Moore},\ and\ \citenamefont {Pollmann}}]{Zaletel2015}%
  \BibitemOpen
  \bibfield  {author} {\bibinfo {author} {\bibfnamefont {M.~P.}\ \bibnamefont
  {Zaletel}}, \bibinfo {author} {\bibfnamefont {R.~S.~K.}\ \bibnamefont
  {Mong}}, \bibinfo {author} {\bibfnamefont {C.}~\bibnamefont {Karrasch}},
  \bibinfo {author} {\bibfnamefont {J.~E.}\ \bibnamefont {Moore}},\ and\
  \bibinfo {author} {\bibfnamefont {F.}~\bibnamefont {Pollmann}},\ }\bibfield
  {title} {\bibinfo {title} {Time-evolving a matrix product state with
  long-ranged interactions},\ }\href
  {https://doi.org/10.1103/PhysRevB.91.165112} {\bibfield  {journal} {\bibinfo
  {journal} {Phys. Rev. B}\ }\textbf {\bibinfo {volume} {91}},\ \bibinfo
  {pages} {165112} (\bibinfo {year} {2015})}\BibitemShut {NoStop}%
\bibitem [{\citenamefont {{Van Damme}}\ \emph {et~al.}(2023)\citenamefont {{Van
  Damme}}, \citenamefont {Haegeman}, \citenamefont {McCulloch},\ and\
  \citenamefont {Vanderstraeten}}]{VanDamme2023}%
  \BibitemOpen
  \bibfield  {author} {\bibinfo {author} {\bibfnamefont {M.}~\bibnamefont {{Van
  Damme}}}, \bibinfo {author} {\bibfnamefont {J.}~\bibnamefont {Haegeman}},
  \bibinfo {author} {\bibfnamefont {I.}~\bibnamefont {McCulloch}},\ and\
  \bibinfo {author} {\bibfnamefont {L.}~\bibnamefont {Vanderstraeten}},\
  }\href@noop {} {\bibinfo {title} {Efficient higher-order matrix product
  operators for time evolution}} (\bibinfo {year} {2023}),\ \Eprint
  {https://arxiv.org/abs/2302.14181} {arXiv:2302.14181} \BibitemShut {NoStop}%
\bibitem [{\citenamefont {Shiba}(1980)}]{Shiba1980}%
  \BibitemOpen
  \bibfield  {author} {\bibinfo {author} {\bibfnamefont {H.}~\bibnamefont
  {Shiba}},\ }\bibfield  {title} {\bibinfo {title} {{Quantization of Magnetic
  Excitation Continuum Due to Interchain Coupling in Nearly One-Dimensional
  Ising-Like Antiferromagnets}},\ }\href {https://doi.org/10.1143/PTP.64.466}
  {\bibfield  {journal} {\bibinfo  {journal} {Progress of Theoretical Physics}\
  }\textbf {\bibinfo {volume} {64}},\ \bibinfo {pages} {466} (\bibinfo {year}
  {1980})}\BibitemShut {NoStop}%
\bibitem [{\citenamefont {Affleck}(1998)}]{Affleck1998}%
  \BibitemOpen
  \bibfield  {author} {\bibinfo {author} {\bibfnamefont {I.}~\bibnamefont
  {Affleck}},\ }\bibinfo {title} {Soliton confinement and the excitation
  spectrum of spin-peierls antiferromagnets},\ in\ \href
  {https://doi.org/10.1007/978-94-011-4988-4_6} {\emph {\bibinfo {booktitle}
  {Dynamical Properties of Unconventional Magnetic Systems}}},\ \bibinfo
  {editor} {edited by\ \bibinfo {editor} {\bibfnamefont {A.~T.}\ \bibnamefont
  {Skjeltorp}}\ and\ \bibinfo {editor} {\bibfnamefont {D.}~\bibnamefont
  {Sherrington}}}\ (\bibinfo  {publisher} {Springer Netherlands},\ \bibinfo
  {address} {Dordrecht},\ \bibinfo {year} {1998})\ pp.\ \bibinfo {pages}
  {123--131}\BibitemShut {NoStop}%
\bibitem [{\citenamefont {Coldea}\ \emph {et~al.}(2010)\citenamefont {Coldea},
  \citenamefont {Tennant}, \citenamefont {Wheeler}, \citenamefont {Wawrzynska},
  \citenamefont {Prabhakaran}, \citenamefont {Telling}, \citenamefont
  {Habicht}, \citenamefont {Smeibidl},\ and\ \citenamefont
  {Kiefer}}]{Coldea2010}%
  \BibitemOpen
  \bibfield  {author} {\bibinfo {author} {\bibfnamefont {R.}~\bibnamefont
  {Coldea}}, \bibinfo {author} {\bibfnamefont {D.~A.}\ \bibnamefont {Tennant}},
  \bibinfo {author} {\bibfnamefont {E.~M.}\ \bibnamefont {Wheeler}}, \bibinfo
  {author} {\bibfnamefont {E.}~\bibnamefont {Wawrzynska}}, \bibinfo {author}
  {\bibfnamefont {D.}~\bibnamefont {Prabhakaran}}, \bibinfo {author}
  {\bibfnamefont {M.}~\bibnamefont {Telling}}, \bibinfo {author} {\bibfnamefont
  {K.}~\bibnamefont {Habicht}}, \bibinfo {author} {\bibfnamefont
  {P.}~\bibnamefont {Smeibidl}},\ and\ \bibinfo {author} {\bibfnamefont
  {K.}~\bibnamefont {Kiefer}},\ }\bibfield  {title} {\bibinfo {title} {Quantum
  criticality in an {I}sing chain: {E}xperimental evidence for emergent {E$_8$}
  symmetry},\ }\href {https://doi.org/10.1126/science.1180085} {\bibfield
  {journal} {\bibinfo  {journal} {Science}\ }\textbf {\bibinfo {volume}
  {327}},\ \bibinfo {pages} {177} (\bibinfo {year} {2010})}\BibitemShut
  {NoStop}%
\bibitem [{\citenamefont {Lake}\ \emph {et~al.}(2010)\citenamefont {Lake},
  \citenamefont {Tsvelik}, \citenamefont {Notbohm}, \citenamefont
  {Alan~Tennant}, \citenamefont {Perring}, \citenamefont {Reehuis},
  \citenamefont {Sekar}, \citenamefont {Krabbes},\ and\ \citenamefont
  {B{\"u}chner}}]{Lake2010}%
  \BibitemOpen
  \bibfield  {author} {\bibinfo {author} {\bibfnamefont {B.}~\bibnamefont
  {Lake}}, \bibinfo {author} {\bibfnamefont {A.~M.}\ \bibnamefont {Tsvelik}},
  \bibinfo {author} {\bibfnamefont {S.}~\bibnamefont {Notbohm}}, \bibinfo
  {author} {\bibfnamefont {D.}~\bibnamefont {Alan~Tennant}}, \bibinfo {author}
  {\bibfnamefont {T.~G.}\ \bibnamefont {Perring}}, \bibinfo {author}
  {\bibfnamefont {M.}~\bibnamefont {Reehuis}}, \bibinfo {author} {\bibfnamefont
  {C.}~\bibnamefont {Sekar}}, \bibinfo {author} {\bibfnamefont
  {G.}~\bibnamefont {Krabbes}},\ and\ \bibinfo {author} {\bibfnamefont
  {B.}~\bibnamefont {B{\"u}chner}},\ }\bibfield  {title} {\bibinfo {title}
  {Confinement of fractional quantum number particles in a condensed-matter
  system},\ }\href {https://doi.org/10.1038/nphys1462} {\bibfield  {journal}
  {\bibinfo  {journal} {Nature Physics}\ }\textbf {\bibinfo {volume} {6}},\
  \bibinfo {pages} {50} (\bibinfo {year} {2010})}\BibitemShut {NoStop}%
\bibitem [{\citenamefont {Bera}\ \emph {et~al.}(2017)\citenamefont {Bera},
  \citenamefont {Lake}, \citenamefont {Essler}, \citenamefont {Vanderstraeten},
  \citenamefont {Hubig}, \citenamefont {Schollw\"ock}, \citenamefont {Islam},
  \citenamefont {Schneidewind},\ and\ \citenamefont
  {Quintero-Castro}}]{Bera2017}%
  \BibitemOpen
  \bibfield  {author} {\bibinfo {author} {\bibfnamefont {A.~K.}\ \bibnamefont
  {Bera}}, \bibinfo {author} {\bibfnamefont {B.}~\bibnamefont {Lake}}, \bibinfo
  {author} {\bibfnamefont {F.~H.~L.}\ \bibnamefont {Essler}}, \bibinfo {author}
  {\bibfnamefont {L.}~\bibnamefont {Vanderstraeten}}, \bibinfo {author}
  {\bibfnamefont {C.}~\bibnamefont {Hubig}}, \bibinfo {author} {\bibfnamefont
  {U.}~\bibnamefont {Schollw\"ock}}, \bibinfo {author} {\bibfnamefont {A.~T.
  M.~N.}\ \bibnamefont {Islam}}, \bibinfo {author} {\bibfnamefont
  {A.}~\bibnamefont {Schneidewind}},\ and\ \bibinfo {author} {\bibfnamefont
  {D.~L.}\ \bibnamefont {Quintero-Castro}},\ }\bibfield  {title} {\bibinfo
  {title} {{Spinon confinement in a quasi-one-dimensional anisotropic
  Heisenberg magnet}},\ }\href {https://doi.org/10.1103/PhysRevB.96.054423}
  {\bibfield  {journal} {\bibinfo  {journal} {Phys. Rev. B}\ }\textbf {\bibinfo
  {volume} {96}},\ \bibinfo {pages} {054423} (\bibinfo {year}
  {2017})}\BibitemShut {NoStop}%
\bibitem [{\citenamefont {Kormos}\ \emph {et~al.}(2017)\citenamefont {Kormos},
  \citenamefont {Collura}, \citenamefont {Tak{\'a}cs},\ and\ \citenamefont
  {Calabrese}}]{Kormos2017}%
  \BibitemOpen
  \bibfield  {author} {\bibinfo {author} {\bibfnamefont {M.}~\bibnamefont
  {Kormos}}, \bibinfo {author} {\bibfnamefont {M.}~\bibnamefont {Collura}},
  \bibinfo {author} {\bibfnamefont {G.}~\bibnamefont {Tak{\'a}cs}},\ and\
  \bibinfo {author} {\bibfnamefont {P.}~\bibnamefont {Calabrese}},\ }\bibfield
  {title} {\bibinfo {title} {Real-time confinement following a quantum quench
  to a non-integrable model},\ }\href {https://doi.org/10.1038/nphys3934}
  {\bibfield  {journal} {\bibinfo  {journal} {Nature Physics}\ }\textbf
  {\bibinfo {volume} {13}},\ \bibinfo {pages} {246} (\bibinfo {year}
  {2017})}\BibitemShut {NoStop}%
\bibitem [{\citenamefont {Cazalilla}\ and\ \citenamefont
  {Rey}(2014)}]{Cazalilla-R-14}%
  \BibitemOpen
  \bibfield  {author} {\bibinfo {author} {\bibfnamefont {M.~A.}\ \bibnamefont
  {Cazalilla}}\ and\ \bibinfo {author} {\bibfnamefont {A.~M.}\ \bibnamefont
  {Rey}},\ }\bibfield  {title} {\bibinfo {title} {Ultracold fermi gases with
  emergent {SU($N$)} symmetry},\ }\href {http://arxiv.org/abs/1403.2792}
  {\bibfield  {journal} {\bibinfo  {journal} {Rep. Prog. Phys.}\ }\textbf
  {\bibinfo {volume} {77}},\ \bibinfo {pages} {124401} (\bibinfo {year}
  {2014})}\BibitemShut {NoStop}%
\bibitem [{\citenamefont {Di~Francesco}\ \emph {et~al.}(1996)\citenamefont
  {Di~Francesco}, \citenamefont {Mathieu},\ and\ \citenamefont
  {S\'{e}n\'{e}chal}}]{DiFrancesco-M-S-book}%
  \BibitemOpen
  \bibfield  {author} {\bibinfo {author} {\bibfnamefont {P.}~\bibnamefont
  {Di~Francesco}}, \bibinfo {author} {\bibfnamefont {P.}~\bibnamefont
  {Mathieu}},\ and\ \bibinfo {author} {\bibfnamefont {D.}~\bibnamefont
  {S\'{e}n\'{e}chal}},\ }\href {https://doi.org/10.1007/978-1-4612-2256-9}
  {\emph {\bibinfo {title} {Conformal Field Theory}}}\ (\bibinfo  {publisher}
  {Springer Verlag},\ \bibinfo {year} {1996})\BibitemShut {NoStop}%
\bibitem [{\citenamefont {Affleck}\ \emph {et~al.}(1989)\citenamefont
  {Affleck}, \citenamefont {Gepner}, \citenamefont {Schulz},\ and\
  \citenamefont {Ziman}}]{Affleck-G-S-Z-89}%
  \BibitemOpen
  \bibfield  {author} {\bibinfo {author} {\bibfnamefont {I.}~\bibnamefont
  {Affleck}}, \bibinfo {author} {\bibfnamefont {D.}~\bibnamefont {Gepner}},
  \bibinfo {author} {\bibfnamefont {H.~J.}\ \bibnamefont {Schulz}},\ and\
  \bibinfo {author} {\bibfnamefont {T.}~\bibnamefont {Ziman}},\ }\bibfield
  {title} {\bibinfo {title} {Critical-behavior of spin-{S} {H}eisenberg
  antiferromagnetic chains - analytic and numerical results},\ }\href
  {https://iopscience.iop.org/article/10.1088/0305-4470/22/5/015} {\bibfield
  {journal} {\bibinfo  {journal} {Journal of Physics A-Mathematical and
  general}\ }\textbf {\bibinfo {volume} {22}},\ \bibinfo {pages} {511}
  (\bibinfo {year} {1989})}\BibitemShut {NoStop}%
\bibitem [{\citenamefont {Majumdar}\ and\ \citenamefont
  {Mukherjee}(2002)}]{Majumdar-M-02}%
  \BibitemOpen
  \bibfield  {author} {\bibinfo {author} {\bibfnamefont {K.}~\bibnamefont
  {Majumdar}}\ and\ \bibinfo {author} {\bibfnamefont {M.}~\bibnamefont
  {Mukherjee}},\ }\bibfield  {title} {\bibinfo {title} {Logarithmic corrections
  to finite-size spectrum of {SU($N$)} symmetric quantum chains},\ }\href
  {http://iopscience.iop.org/0305-4470/35/38/101/fulltext/} {\bibfield
  {journal} {\bibinfo  {journal} {J. Phys. A: Math. Gen.}\ }\textbf {\bibinfo
  {volume} {35}},\ \bibinfo {pages} {L543} (\bibinfo {year}
  {2002})}\BibitemShut {NoStop}%
\bibitem [{\citenamefont {Nonne}\ \emph {et~al.}(2013)\citenamefont {Nonne},
  \citenamefont {Moliner}, \citenamefont {Capponi}, \citenamefont
  {Lecheminant},\ and\ \citenamefont {Totsuka}}]{Nonne2013}%
  \BibitemOpen
  \bibfield  {author} {\bibinfo {author} {\bibfnamefont {H.}~\bibnamefont
  {Nonne}}, \bibinfo {author} {\bibfnamefont {M.}~\bibnamefont {Moliner}},
  \bibinfo {author} {\bibfnamefont {S.}~\bibnamefont {Capponi}}, \bibinfo
  {author} {\bibfnamefont {P.}~\bibnamefont {Lecheminant}},\ and\ \bibinfo
  {author} {\bibfnamefont {K.}~\bibnamefont {Totsuka}},\ }\bibfield  {title}
  {\bibinfo {title} {Symmetry-protected topological phases of alkaline-earth
  cold fermionic atoms in one dimension},\ }\href
  {http://stacks.iop.org/0295-5075/102/i=3/a=37008} {\bibfield  {journal}
  {\bibinfo  {journal} {Europhys. Lett.}\ }\textbf {\bibinfo {volume} {102}},\
  \bibinfo {pages} {37008} (\bibinfo {year} {2013})}\BibitemShut {NoStop}%
\bibitem [{\citenamefont {Bois}\ \emph {et~al.}(2015)\citenamefont {Bois},
  \citenamefont {Capponi}, \citenamefont {Lecheminant}, \citenamefont
  {Moliner},\ and\ \citenamefont {Totsuka}}]{Bois2015}%
  \BibitemOpen
  \bibfield  {author} {\bibinfo {author} {\bibfnamefont {V.}~\bibnamefont
  {Bois}}, \bibinfo {author} {\bibfnamefont {S.}~\bibnamefont {Capponi}},
  \bibinfo {author} {\bibfnamefont {P.}~\bibnamefont {Lecheminant}}, \bibinfo
  {author} {\bibfnamefont {M.}~\bibnamefont {Moliner}},\ and\ \bibinfo {author}
  {\bibfnamefont {K.}~\bibnamefont {Totsuka}},\ }\bibfield  {title} {\bibinfo
  {title} {Phase diagrams of one-dimensional half-filled two-orbital
  $\mathrm{SU}(n)$ cold fermion systems},\ }\href
  {https://doi.org/10.1103/PhysRevB.91.075121} {\bibfield  {journal} {\bibinfo
  {journal} {Phys. Rev. B}\ }\textbf {\bibinfo {volume} {91}},\ \bibinfo
  {pages} {075121} (\bibinfo {year} {2015})}\BibitemShut {NoStop}%
\bibitem [{\citenamefont {Papenbrock}\ \emph {et~al.}(2003)\citenamefont
  {Papenbrock}, \citenamefont {Barnes}, \citenamefont {Dean}, \citenamefont
  {Stoitsov},\ and\ \citenamefont {Strayer}}]{Papenbrock2003}%
  \BibitemOpen
  \bibfield  {author} {\bibinfo {author} {\bibfnamefont {T.}~\bibnamefont
  {Papenbrock}}, \bibinfo {author} {\bibfnamefont {T.}~\bibnamefont {Barnes}},
  \bibinfo {author} {\bibfnamefont {D.~J.}\ \bibnamefont {Dean}}, \bibinfo
  {author} {\bibfnamefont {M.~V.}\ \bibnamefont {Stoitsov}},\ and\ \bibinfo
  {author} {\bibfnamefont {M.~R.}\ \bibnamefont {Strayer}},\ }\bibfield
  {title} {\bibinfo {title} {{Density matrix renormalization group study of
  critical behavior of the
  $\mathrm{s}\mathrm{p}\mathrm{i}\mathrm{n}\ensuremath{-}\frac{1}{2}$
  alternating Heisenberg chain}},\ }\href
  {https://doi.org/10.1103/PhysRevB.68.024416} {\bibfield  {journal} {\bibinfo
  {journal} {Phys. Rev. B}\ }\textbf {\bibinfo {volume} {68}},\ \bibinfo
  {pages} {024416} (\bibinfo {year} {2003})}\BibitemShut {NoStop}%
\bibitem [{\citenamefont {Orignac}(2004)}]{Orignac2004}%
  \BibitemOpen
  \bibfield  {author} {\bibinfo {author} {\bibfnamefont {E.}~\bibnamefont
  {Orignac}},\ }\bibfield  {title} {\bibinfo {title} {Quantitative expression
  of the spin gap via bosonization for a dimerized spin-1/2 chain},\ }\href
  {https://doi.org/10.1140/epjb/e2004-00198-5} {\bibfield  {journal} {\bibinfo
  {journal} {Eur. Phys. J. B}\ }\textbf {\bibinfo {volume} {39}},\ \bibinfo
  {pages} {335} (\bibinfo {year} {2004})}\BibitemShut {NoStop}%
\bibitem [{\citenamefont {S\o{}rensen}\ and\ \citenamefont
  {Young}(1990)}]{Sorensen1990}%
  \BibitemOpen
  \bibfield  {author} {\bibinfo {author} {\bibfnamefont {E.~S.}\ \bibnamefont
  {S\o{}rensen}}\ and\ \bibinfo {author} {\bibfnamefont {A.~P.}\ \bibnamefont
  {Young}},\ }\bibfield  {title} {\bibinfo {title} {Correlation length of the
  biquadratic spin-1 chain},\ }\href {https://doi.org/10.1103/PhysRevB.42.754}
  {\bibfield  {journal} {\bibinfo  {journal} {Phys. Rev. B}\ }\textbf {\bibinfo
  {volume} {42}},\ \bibinfo {pages} {754} (\bibinfo {year} {1990})}\BibitemShut
  {NoStop}%
\bibitem [{\citenamefont {Baxter}(1982)}]{Baxter_book}%
  \BibitemOpen
  \bibfield  {author} {\bibinfo {author} {\bibfnamefont {R.~J.}\ \bibnamefont
  {Baxter}},\ }\href {https://doi.org/10.1142/9789814415255_0002} {\emph
  {\bibinfo {title} {{Exactly solved models in statistical mechanics}}}}\
  (\bibinfo  {publisher} {Academic Press},\ \bibinfo {year} {1982})\BibitemShut
  {NoStop}%
\end{thebibliography}
\end{document}